\documentclass[aps, pra,  twocolumn, superscriptaddress, amsmath]{revtex4-2}

\usepackage{graphicx}       %
\usepackage{soul}
\usepackage[colorlinks, breaklinks, linkcolor=red,  citecolor=red, urlcolor=red]{hyperref}
\newcommand{\qqq}[1]{#1}
\usepackage{comment}

\renewcommand{\Im}{\mathop{\mathrm{Im}}\nolimits}
\renewcommand{\Re}{\mathop{\mathrm{Re}}\nolimits}
\newcommand{\Tr}{\mathop{\mathrm{Tr}}\nolimits}

\begin{document}
\title{Theory of the photonic Joule effect in superconducting circuits
}

\author{Samuel Cailleaux}
    \affiliation{Univ. Grenoble Alpes, CNRS, Grenoble INP, Institut N\'eel, 38000 Grenoble, France}
    \affiliation{Univ. Grenoble Alpes, CNRS, LPMMC, 38000 Grenoble, France}

\author{Quentin Ficheux}
    \affiliation{Univ. Grenoble Alpes, CNRS, Grenoble INP, Institut N\'eel, 38000 Grenoble, France}
	
\author{Nicolas Roch}
    \affiliation{Univ. Grenoble Alpes, CNRS, Grenoble INP, Institut N\'eel, 38000 Grenoble, France}

\author{Denis M. Basko}
    \affiliation{Univ. Grenoble Alpes, CNRS, LPMMC, 38000 Grenoble, France}

\date{\today}

\begin{abstract}

When a small system is coupled to a bath, it is generally assumed that the state of the bath remains unaffected by the system due to the bath's large number of degrees of freedom. Here we show theoretically that this assumption can be easily violated for photonic baths typically used in experiments involving superconducting circuits.
We analyze the dynamics of a voltage-biased Josephson junction coupled to a photonic bath, represented as a long Josephson junction chain.
Our findings show that the system can reach a non-equilibrium steady state where the photonic degrees of freedom become significantly overheated, leading to a qualitative change in the current-voltage $I-V$  curve.
This phenomenon is analogous to the Joule effect observed in electrical conductors, where flowing current can substantially heat up electrons. Recognizing this effect is crucial for the many applications of high-impedance environments in quantum technologies. 

\end{abstract}

\maketitle

\textit{Introduction.---}
Controlling and engineering dissipation open broad perspectives in quantum science and technology~\cite{harrington_engineered_2022}. Superconducting quantum circuits, whose photonic degrees of freedom \qqq{can mimic the Caldeira-Leggett model of a bath~\cite{caldeira_quantum_1983}}, offer a promising platform for making synthetic dissipative environments with engineered properties. 
In contrast to conventional solid-state resistors, synthetic photonic baths also provide a possibility to monitor their internal state and to characterize the energy deposited within them~\cite{gabelli_hanbury_2004, hofheinz_bright_2011, fraudet_direct_2024}.
Such engineered environments have been successfully used to simulate paradigmatic models of quantum electrodynamics and many-body physics in regimes, difficult to access on other platforms~\cite{sundaresan_beyond_2015, Jaako2016, forndiaz_ultrastrong_2017, puertasmartinez_tunable_2019, kuzmin_superstrong_2019, kuzmin_inelastic_2021, leger_revealing_2023}. 
Understanding their properties is crucial for many applications~\cite{Crescini2023, corlevi2006phase, giacomelli2024emergent, Castellanos-Beltran2008, Esposito2021, kitaev2006protectedqubitbasedsuperconducting, Nataf2011, Liu2017, Mirhosseini2018, Sinha2022, Zhang2023}.

\begin{figure}
    \centering
    \includegraphics[width=0.47\textwidth]{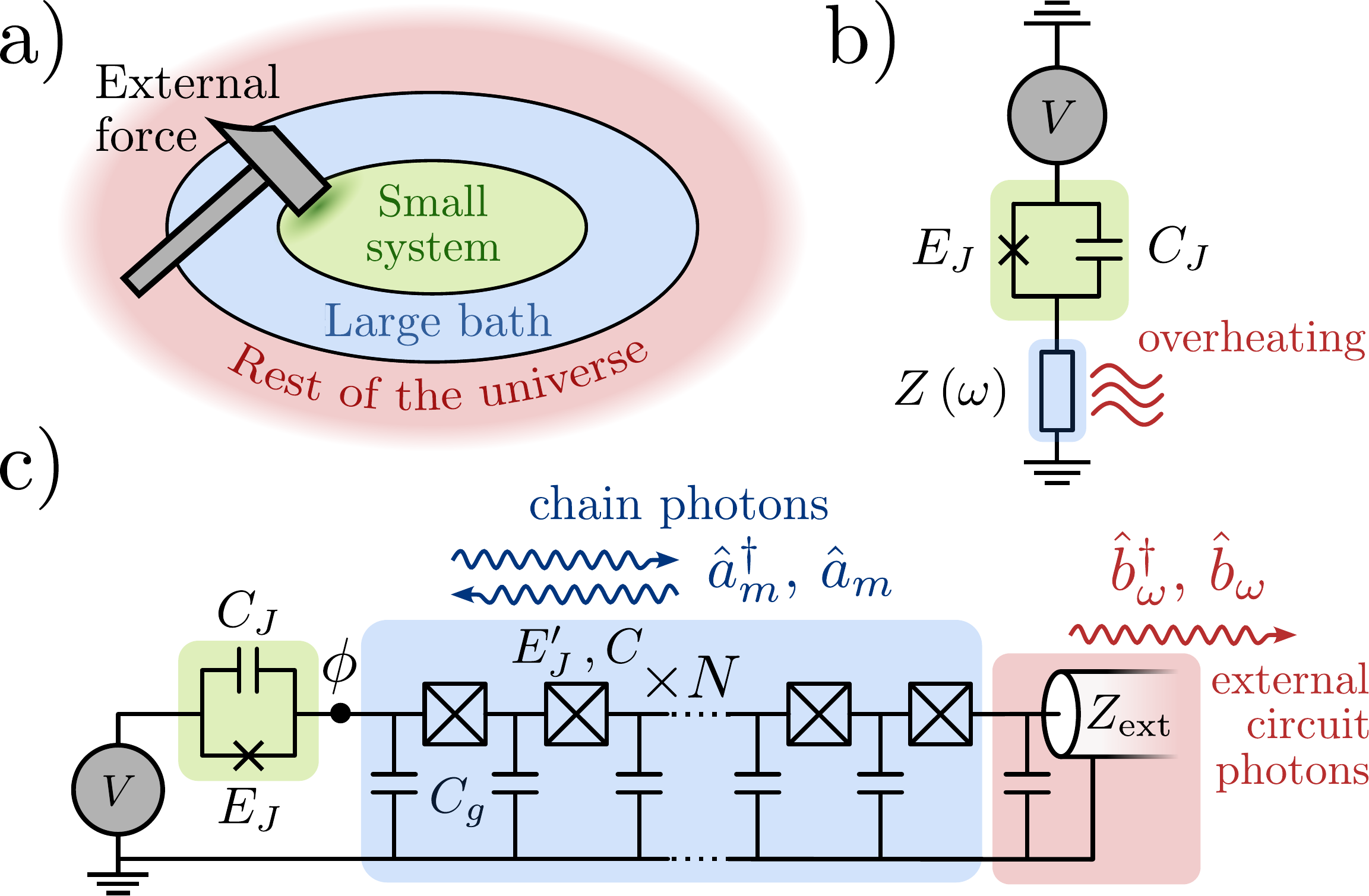}
    \caption{\qqq{(a)~The open-system paradigm: a small subsystem is manipulated by an external force and coupled to a large bath. The bath is coupled to the external world, which is often disregarded when the bath's state is assumed to remain unchanged.}
    \qqq{(b)~Inelastic Cooper pair tunneling circuit: a small JJ with Josephson energy $E_J$ and capacitance~$C_J$ (green), subject to a dc voltage bias $V$, coupled to a dissipative impedance $Z(\omega)$ (blue). The impedance is overheated when the Joule heating power exceeds its cooling power.
    (c)~The impedance implementated as a long chain of $N$ large JJs with the Josephson energies $E_J' \gg E_J$, capacitances $C$, and ground capacitances $C_\text{g}$. The chain is coupled to the external circuit (red), represented by a transmission line of impedance $Z_\text{ext}=50\:\Omega$, which cools the chain by photon escape.}
    }
    \label{fig:circuit}
\end{figure}

\qqq{
A physical system is expected to behave effectively as a bath, when it (i)~has a sufficiently broad and dense spectrum of excitations to forget initial conditions, and (ii)~is sufficiently large to remain unaffected by the small subsystem it is coupled to.
Here we provide a counter-example to this common expectation.
We consider one small Josephson junction (JJ) coupled to a chain of several thousands large junctions, which seemingly meets these conditions (Fig.~\ref{fig:circuit}).
However, the photonic excitations of the chain are very long-lived. 
}
Indeed, at low temperatures and subgap frequencies, there are very few quasiparticles, and photonic degrees of freedom are not directly coupled to phonons. The photons are mainly damped by escape into the external circuit, which occurs at the boundary %
and is typically suppressed by impedance mismatch. %
If a bath, no matter how large, evacuates energy mainly at the boundary, the energy may easily accumulate inside. Theoretical investigation of this possibility is the subject of the present paper.

We study the paradigmatic situation of inelastic Cooper pair tunneling in a voltage-biased small JJ coupled to a dissipative harmonic environment. 
When tunneling across the small junction, a Cooper pair must exchange energy with the environment; this process determines the dc current through the junction. The standard description of this phenomenon via the so-called $P(E)$~theory~\cite{ingold_charge_2005} assumes the environment to be in a given thermal state.
We relax this assumption and show that in a realistic superconducting circuit the environment's photons tend to accumulate and induce a strong back-action on the inelastic tunneling, leading to a non-equilibrium steady state of the whole system and a dramatic modification of the $I-V$ curve.
The system may even enter a bistable regime with a hysteretic $I-V$ curve, somewhat analogous to bistabilities arising in electronic transport when the Joule heat deposited in the electronic subsystem is not evacuated into phonons quickly enough~\cite{gurevich_selfheating_1987}.
Thus, the situation studied here can be viewed as a photonic analog of the familiar Joule effect in electrical conductors.

We analyze the environment's internal state using several complementary approaches.
First, assuming the dynamics of the environments's modes to be fully incoherent, we generalize the standard $P(E)$ theory to the case of arbitrary modes' occupations, which are themselves determined self-consistently from a kinetic equation.
This calculation is valid only for a sufficiently dense frequency spectrum of the modes; in order to control the effect of mode discreteness,  we do a complementary calculation solving the classical equations of motion for the modes' amplitudes.
These approaches agree in a large interval of voltages due to the chaotic character of the modes' classical dynamics, which generates an effective noise destroying the coherence between the modes.
In the resulting non-equilibrium steady state, each mode is thermal, but different modes may have different temperatures.

\textit{Model.---}
We consider a small JJ with Josephson energy $E_J$ and capacitance~$C_J$, subject to a dc voltage bias~$V$, and coupled to a 
\qqq{long chain of $N$~large JJs characterized by the Josephson energy $E_J'\gg{E}_J$, junction capacitance~$C$, and capacitance to ground~$C_\text{g}\ll{C}$. We assume $E_J'\gg{e}^2/C$ ($e>0$ being the elementary charge), so the chain is in the harmonic regime. An infinite chain would (i)~host plane wave photonic modes with frequency dispersion
$\omega(k)=\omega_\text{p}k(k^2+C_\text{g}/C)^{-1/2}$ as a function of the dimensionless wave vector~$k$~\cite{masluk_microwave_2012}, and (ii)~behave as a linear circuit element with the impedance $Z(\omega)=Z_0(1-\omega^2/\omega_\text{p}^2)^{-1/2}$, where $Z_0^{-1}=(2e/\hbar)\sqrt{E_J'C_\text{g}}$. 
The chain's plasma frequency $\omega_\text{p}\equiv(2e/\hbar)\sqrt{E_J'/C}$ coincides with that of a single large junction (in the limit $C_\text{g}\ll{C}$), and provides an upper bound to $\omega(k)$ which translates into a sharp high-frequency cutoff in $\Re{Z}(\omega)$.
}
The shunting effect of the small junction's capacitance~$C_J$ on the total impedance seen by the junction, $Z_\text{tot}(\omega)=1/[1/Z(\omega)-i\omega{C}_J]$, provides an additional soft cutoff at frequencies $\omega\sim1/(Z_0C_J)$. 
We assume the two to be of the same order (typically, in the range 10--20~GHz), well below the superconducting gap of the material ($2\Delta\approx 2\pi\hbar\times100\:\mbox{GHz}$ for aluminum).

\begin{figure}
    \centering
    \includegraphics[width=0.48\textwidth]{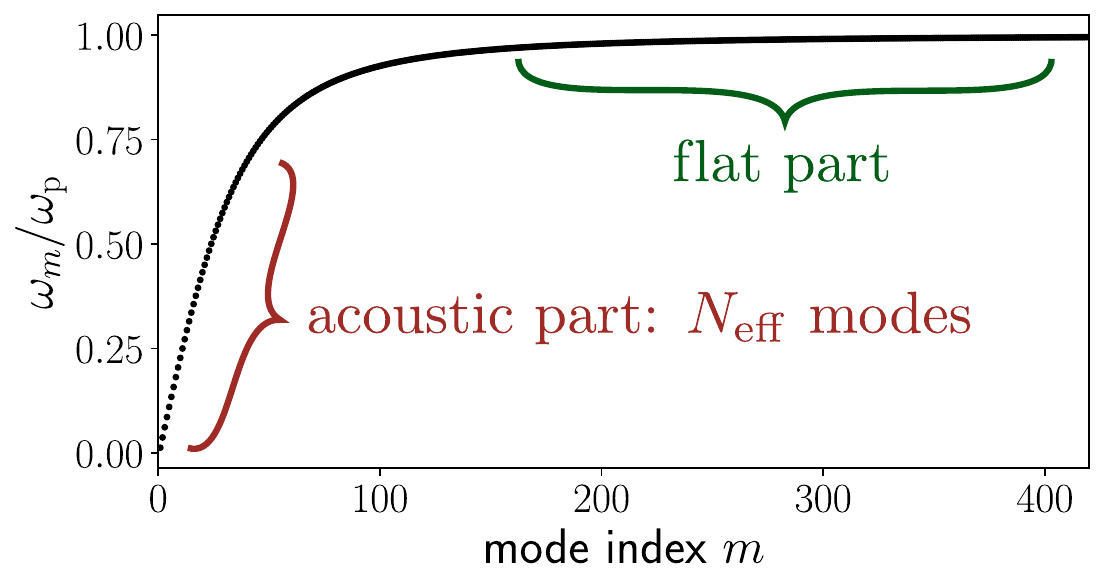}
    \caption{Mode frequencies $\{\omega_m\}$ for a chain of $N=5000$ junctions with typical parameters: $Z_0=5\,\mathrm{k\Omega}$, $\omega_\text{p}=2\pi\times20\:\mbox{GHz}$, $C/C_\text{g}=1600$. 
    The frequency spacing on the acoustic part of the dispersion is $\delta_1=(\pi/N)\sqrt{C/C_g}\,\omega_\text{p}=2\pi\times0.5\:\mbox{GHz}$, yielding $N_\text{eff}=\omega_\text{p}/\delta_1=40$.
    }
    \label{fig:dispersion_circled}
\end{figure}

\qqq{
In reality, a chain with a finite number $N\gg1$ of junctions is connected on one end to the small junction, and on the other to the external circuit which is modeled as a shunting resistance $Z_\text{ext}\ll{Z}_0$ (Fig.~\ref{fig:circuit}). 
Such boundary conditions (i)~effectively quantize the wave vectors $k_m$, $m=1,2,\ldots$, producing discrete Fabry-Perot-like modes with frequencies $\omega_m=\omega(k_m)$ (Fig.~\ref{fig:dispersion_circled}), and (ii) give each mode~$m$ a small decay rate~$\kappa_m\ll\delta_m\equiv\omega_{m+1}-\omega_m$, the mode frequency spacing~(see Appendix~\ref{app:param_computation} for details). Note that $\delta_m$ varies with~$m$ due to the curvature of $\omega(k)$.
}

This system can be described by the following Hamiltonian~\cite{armour_universal_2013, armour_josephson_2015, trif_photon_2015, hofer_quantum_2016, dambach_generating_2017, cassidy_demonstration_2017, simon_theory_2018, aiello2020thesis, lang_discrete_2023}:
\begin{align}
    \hat{H} = {}&{} {-E_J}\cos\left[
    (2e/\hbar)Vt-\hat\phi
    \right]
    + \sum_m\hbar\omega_m\hat{a}_m^\dagger\hat{a}_m \nonumber\\
    {}&{} + \int\limits_0^\infty{d\omega}\left[
    \hbar\omega\,\hat{b}^\dagger_\omega\hat{b}_\omega
    + \sum_m\hbar\sqrt{\frac{\kappa_m}{2\pi}}
    \left(\hat{a}_m^\dagger\hat{b}_\omega + \hat{b}_\omega^\dagger\hat{a}_m\right)\right].
    \label{eq:Hamiltonian}
\end{align}
\qqq{The phase drop $\hat\phi=\sum_m\Lambda_m(\hat{a}_m+\hat{a}_m^\dagger)$ over the chain is expressed via the dimensionless coupling constants~$\Lambda_m$ and the bosonic creation/annihilation operators $\hat{a}_m^\dagger,\hat{a}_m$ of the chain's photons. These are coupled to the external circuit, whose modes are described by the bosonic operators~$\hat{b}^\dagger_\omega,\hat{b}_\omega$, as expressed by the second line in Eq.~(\ref{eq:Hamiltonian}).
}

\qqq{
The coupling constants $\Lambda_m$ are determined by the mode amplitudes at the first island of the chain. They can be related to the impedance $\mathcal{Z}_\text{tot}(\omega)$ of the capacitively shunted finite-length chain as~\cite{ingold_charge_2005}
\begin{equation}\label{eq:Z=Lambdak}
\frac{(2e)^2}{\hbar}\,\frac{\mathcal{Z}_\text{tot}(\omega)}{i\omega}=
\sum_m\frac{2\omega_m\Lambda_m^2}{(\omega+i\kappa_m/2)^2-\omega_m^2}.
\end{equation}
When $\kappa_m\ll\delta_m$, $\mathcal{Z}_\text{tot}(\omega)$ is given by a dense sequence of sharp peaks. When smoothed over a small frequency interval containing several peaks, it should cross over to $Z_\text{tot}(\omega)$ of the infinite chain. Thus, the coupling constants must be given by
\begin{equation}\label{eq:Lambdam2=}
    \Lambda_m^2 =\frac{(2e)^2\delta_m}{\pi\hbar\omega_m}\Re{Z_\text{tot}(\omega_m)}.
\end{equation}
Note that both $\kappa_m,\delta_m\propto1/N$, while $\Lambda_m\propto1/\sqrt{N}$.
}

\textit{Single-temperature ansatz.---}
For a simple qualitative estimate, we note that for a long JJ chain, only modes in the acoustic part of the dispersion are effectively coupled to the small junction (Fig.~\ref{fig:dispersion_circled}), while most modes have $\omega_m\to\omega_\text{p}$, 
\qqq{vanishing spacing $\delta_m$, and thus vanishing coupling  according to Eq.~(\ref{eq:Lambdam2=}) (see Appendix~\ref{app:param_computation}).}
Then, we make a simplifying assumption that there are \qqq{$N_\text{eff}=\omega_\text{p}/\delta_1$} coupled modes \qqq{with} the same decay rate~$\kappa$ and the same temperature~$T\gtrsim\hbar\omega_\text{p}$  (we set the Boltzmann constant $k_\text{B}=1$), \qqq{to} be found from the energy balance condition:
\begin{equation}
    N_\text{eff}\frac{dT}{dt} =  - \kappa N_\text{eff} T + I( V, T) V. \label{eq:eom_temperature}
\end{equation}
In the steady state, the power injected from the voltage source, $IV$, should match the power evacuated into the external circuit, $\sim\kappa{N}_\text{eff}T$. The dc current~$I_\text{dc}$ is due to inelastic tunneling of Cooper pairs across the small junction, when each tunneling event is accompanied by an energy exchange of $\pm2eV$ with the environment. A perturbative (in $E_J$) calculation of the tunneling rate using the Fermi Golden Rule and assuming a thermal environment, also known as $P(E)$~theory~\cite{ingold_charge_2005}, gives
\begin{subequations}
\label{eqs:PofE}
\begin{align}
&    I_\text{dc} = \frac{\pi e E_J^2}{\hbar} \left[ P(2eV) - P(-2eV) \right],
\label{eq:IVPofE}\\
&    P(E) = \int^{+\infty}_{-\infty} e^{iEt/\hbar+J(t)}\frac{dt}{2\pi\hbar}.
\label{eq:PofE}
\end{align}
$J(t)$ is the phase correlation function of the environment,
\begin{align}
J(t) = {}&{} \int^{+\infty}_0 \frac{d\omega}{\pi}\,\frac{(2e)^2}{\hbar\omega}\Re{Z_\text{tot}(\omega)} \nonumber\\
& \times   \left\{\left[\bar{n}(\omega) + 1\right]e^{-i\omega{t}} + \bar{n}(\omega){e}^{i\omega{t}} - [2\bar{n}(\omega) + 1] \right\},
\label{eq:Joft}
\end{align}
\end{subequations}
determined by the bosonic occupations of the modes, $\bar{n}(\omega)=1/(e^{\hbar\omega/T}-1)$.
This approach has been used to describe the dynamics of a JJ coupled to a resistor made of a conventional metal~\cite{subero_bolometric_2023}, where the weak overheating of the resistor's electrons resulted in a small correction to the $I-V$ curve. In contrast, overheating of photonic degrees of freedom can be strong, changing the $I-V$ curve qualitatively, as we show below.

\begin{figure}
    \centering
    \includegraphics[width=0.49\textwidth]{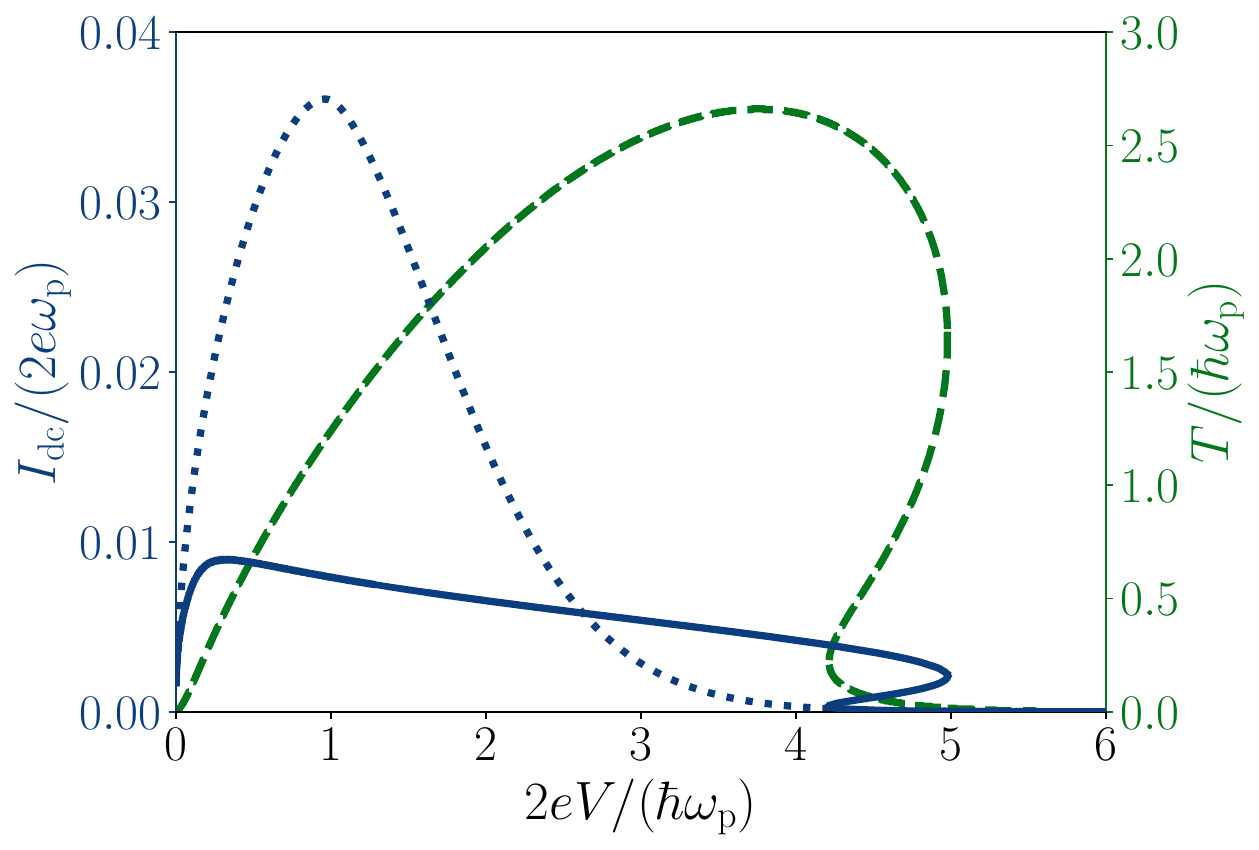}
    \caption{Steady-state dc current (solid line, left axis) and temperature (dashed line, right axis) versus voltage, as found from Eq.~(\ref{eq:eom_temperature}) for %
    the same JJ chain as in Fig.~\ref{fig:dispersion_circled}.
    Coupling to the external circuit with impedance $Z_\text{ext}=50\:\Omega$ gives the decay rate $\kappa=(2/\pi)(Z_\text{ext}/Z_0)\delta_m=2\times10^{4}\:\mbox{s}^{-1}$. The small junction has $E_J=0.2\,\hbar\omega_\text{p}$, $C_J = 2\:\mathrm{fF}$.
    For comparison, we show the zero-temperature $I-V$ curve (dotted line, left axis).
    }
    \label{fig:selfconsistent_temp}
\end{figure}

In Fig.~\ref{fig:selfconsistent_temp} we plot the steady-state dc current and temperature found numerically from Eq.~(\ref{eq:eom_temperature}) for an environment represented by a long JJ chain with typical parameters.
We see that the photonic temperature may well exceed $\hbar\omega_\text{p}\sim1\:\mbox{K}$ and even the aluminum critical temperature $T_c=1.2\:\mbox{K}$.
\qqq{Even then, the superconductivity in the material is not destroyed because the photons are effectively decoupled from quasiparticles (see Appendix~\ref{app:quasiparticles} for an estimate based on Ref.~\cite{catelani_nonequilibrium_2019}).}
The $I-V$ curve of the overheated environment differs significantly from that at $T=0$ (or at $T\sim10-20\:\mbox{mK}$, which would look the same on the scale of the figure). 

In an interval of voltages well above $\hbar\omega_\text{p}/(2e)$, the system even exhibits two stable states with strongly different temperatures.
The low-temperature solution is stable since dissipating the energy $2eV$ requires the emission of many photons which has a small matrix element; consequently, Cooper pairs do not tunnel efficiently, resulting in a small current flowing across the junction. In contrast, at high temperatures, the $P(E)$ function extends well beyond $\hbar\omega_\text{p}$ as (i)~many multiphoton states are available for energy exchange, and (ii)~the tunneling matrix element is large due to the bosonic enhancement; this results in a large current which heats up the environment and stabilizes the high-temperature state.

\textit{$P(E)$ theory with kinetic equation.}
For a quantitative study, we must relax the assumption of all modes having the same temperature. Indeed, the mode populations can thermalize only due to some nonlinear coupling, and the main source of nonlinearity is the small junction itself, described by the first term in Eq.~(\ref{eq:Hamiltonian}). That is, a Cooper pair tunneling across the small junction is accompanied by a redistribution of the photon populations between different modes. This process corresponds to the absorption and emission of energy quanta of $2eV$, described by the time-dependent Hamiltonian~(\ref{eq:Hamiltonian}).

\qqq{
Instead of assuming the average mode occupations $\langle{n}_m\rangle$ to be thermal as in the standard $P(E)$ theory, we determine them from a kinetic equation. It is obtained by using the Fermi Golden Rule for the transition from a Fock state with mode occupations $\{n_m\}$ to another Fock state with occupations $\{n_m'\}$ (akin to Ref.~\cite{hofheinz_bright_2011}), and writing rate equations in the photonic Fock space (see Appendix~\ref{app:kinetic} for  details).
}
Assuming $\langle{n}_m\rangle=\bar{n}(\omega_m)$, $\kappa_m=\kappa(\omega_m)$, and $1/\delta_m=\nu(\omega_m)$, to be smooth functions of $\omega_m$, and taking $Z_\text{tot}(\omega)$ for the infinite system, we write the kinetic equation as
\begin{subequations}\label{eqs:kinetic}\begin{align}
\label{eq:kinetic}
&\hbar\omega\,\nu(\omega)\left[\frac{\partial}{\partial{t}}+\kappa(\omega)\right]\bar{n}(\omega) = W_+(\omega)[\bar{n}(\omega)+1] \nonumber\\
& \hspace*{4.5cm} {} -W_-(\omega)\bar{n}(\omega),\\
&W_\pm(\omega)\equiv\frac{E_J^2}{2}\frac{(2e)^2}{\hbar}\Re{Z_\text{tot}(\omega)}
\sum_{\sigma=\pm}P(2eV\sigma\mp\hbar\omega).
\end{align}\end{subequations}
where $P(E)$ is a functional of $\bar{n}(\omega)$, defined by Eqs.~(\ref{eq:PofE}),~(\ref{eq:Joft}). This gives a closed system of self-consistent equations for $\bar{n}(\omega)$.
Since $\nu(\omega)\propto{N}$, $\kappa(\omega)\propto1/N$, while all other quantities in Eq.~(\ref{eq:kinetic}) do not scale with~$N$, the stationary solution for $\bar{n}(\omega)$ does not scale with~$N$, but the relaxation time \qqq{$\tau\propto{N}$. The experimentally relevant order of limits is $t\to\infty$ first, then $N\to\infty$, unlike the standard one in statistical physics.}

\begin{figure}
    \centering
    \includegraphics[width=0.48\textwidth]{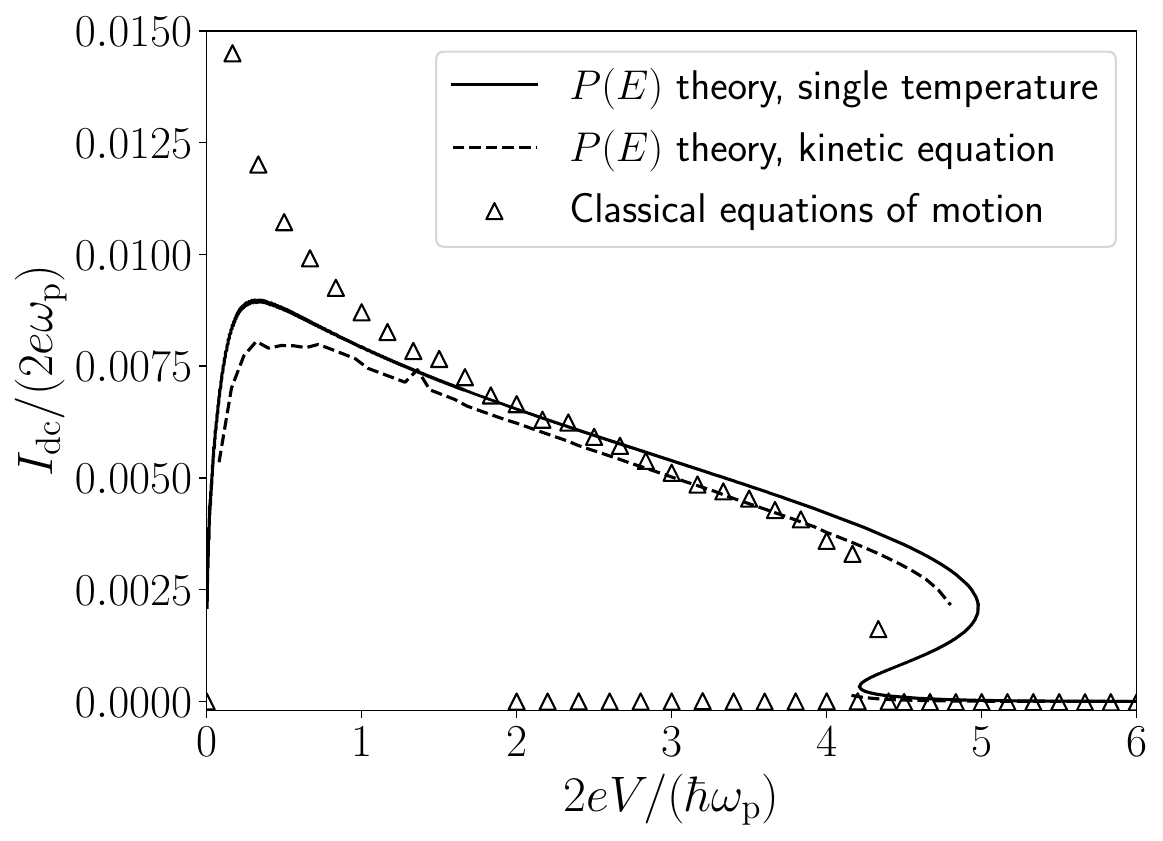}
    \caption{$I-V$ curves found by three different calculations (see the inset) for the same system as in Fig.~\ref{fig:selfconsistent_temp}. The sets $\{\omega_m\},\,\{\kappa_m\},\,\{\Lambda_m\}$ are obtained for the JJ chain shunted by the capacitance~$C_J$ and the resistance $Z_\text{ext}$ on the respective ends (see Appendix~\ref{app:param_computation} for details). A non-radiative damping $\kappa_m^\text{nr}=\omega_m/Q_\text{int}$ with the internal quality factor $Q_\text{int}=5\times10^4$ was added phenomenologically. \qqq{The classical calculation was done with 200 modes.}}
    \label{fig:IVcurves}
\end{figure}

\begin{figure}
    \centering
    \includegraphics[width=0.465\textwidth]{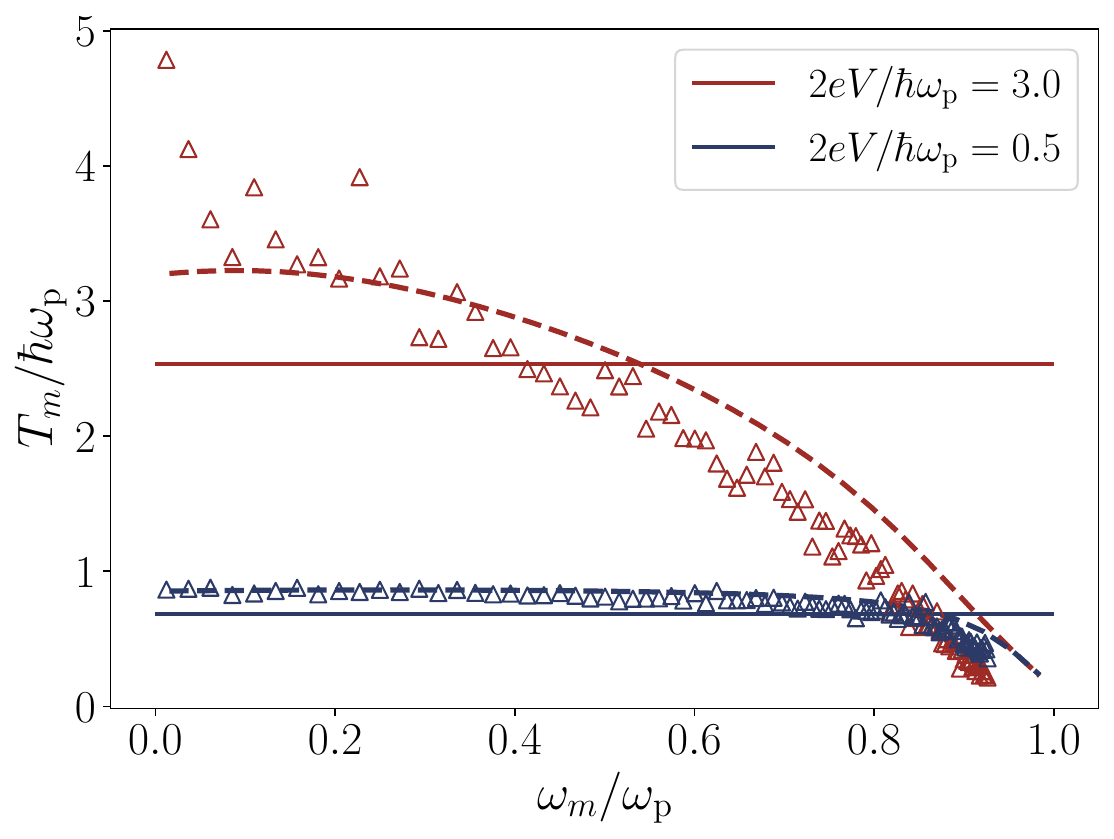}
    \caption{Mode temperatures for the same system as in Figs.~\ref{fig:selfconsistent_temp} and~\ref{fig:IVcurves} at two values of the voltage bias, $2eV/(\hbar\omega_\text{p}) = 0.5,\,3.0$ (blue and red colors, respectively), as found from three different calculations: (i)~single-temperature ansatz (horizontal solid lines), (ii)~kinetic equation with Bose-Einstein $T_m=\hbar\omega_m/\ln[1+1/\bar{n}(\omega_m)]$ (dashed lines), and (iii)~classical equations of motion with $T_m=\hbar\omega_m\,\overline{|\alpha_m(t)|^2}$ from the classical Rayleigh–Jeans law (empty triangles).
    }
    \label{fig:temperatures}
\end{figure}

The $I-V$ curve is determined from $\bar{n}(\omega)$ via Eq.~(\ref{eq:IVPofE}), and is shown by the dashed line in Fig.~\ref{fig:IVcurves} for the same parameters as in Fig.~\ref{fig:selfconsistent_temp}. It matches surprisingly well the single-temperature calculation,
even though 
the resulting mode temperatures $T_m=T(\omega_m)$ are, generally speaking, different for different modes, as shown in Fig.~\ref{fig:temperatures}. 

\qqq{
Perturbative in $E_J$, Eq.~(\ref{eq:kinetic}) requires $E_J\ll\hbar\omega_\text{p},2eV$; on the other hand, for $E_J^2\ll{N}_\text{eff}\hbar^2\kappa\omega_\text{p}$ the heating effect disappears, as seen from Eq.~(\ref{eq:eom_temperature}) at $2eV\sim{T}\sim\hbar\omega_\text{p}$ with $P(E)\sim(\hbar\omega_\text{p})^{-1}$.
Replacement of the finite-$N$ impedance $\mathcal{Z}_\text{tot}(\omega)$ with $Z_\text{tot}(\omega)$ is valid as long as they give the same smooth $P(E)$, which holds for $T\gtrsim0.1\,\hbar\omega_\text{p}$ (see Appendix~\ref{app:kinetic}). The smooth $P(E)$, which is due to multiphoton transitions, validates the Golden Rule calculation in our system with weakly broadened discrete single-photon levels.
Equation~(\ref{eq:kinetic}) also relies on the assumption $\Lambda_m\ll 1$, allowing to neglect inter-mode correlations and to treat the Josephson term in a mean-field fashion so that each mode sees others as a dissipative bath. Seemingly weak (since $\Lambda_m\propto{N}^{-1/2}$), this assumption may break down for a few lowest modes with $\omega_m=(2m-1)\omega_1$. Indeed, Eq.~(\ref{eq:Lambdam2=}) then gives $\Lambda_m=\sqrt{2/[g(m-1/2)]}$ with $g\equiv2\pi\hbar/[(2e)^2Z_0]$ being the dimensionless admittance. Typically, $g\sim1$, so $\Lambda_{m\sim1}\sim1$.
}

\textit{Classical equations of motion.---}
\qqq{
To check the effect of large $\Lambda_m$, we treat the system classically, which is justified when $n_m\sim{T}_m/(\hbar\omega_m)\gg1$. For our parameters, this holds for most modes when $2eV\gtrsim\hbar\omega_\text{p}$, as seen in Fig.~\ref{fig:temperatures}. It will hold even better for larger $E_J$ and~$g$, so the classical calculation is complementary to $P(E)$ theory, with overlapping regions of validity.
}

\qqq{
From the Hamiltonian~(\ref{eq:Hamiltonian}) we obtain the Heisenberg equations of motion for the mode operators $\hat{a}_m(t)$, we eliminate the operators~$\hat{b}_\omega$, which produces damping terms, and we take the average, replacing $\langle\sin(2eVt/\hbar-\hat\phi)\rangle\to\sin(2eVt/\hbar-\langle\hat\phi\rangle)$. This yields a closed system of equations for complex variables $\alpha_m(t)\equiv\langle\hat{a}_m(t)\rangle$:
\begin{subequations}\label{eqs:eom_modes}\begin{align}
& \frac{d\alpha_m}{dt} = \left(-i\omega_m -\frac{\kappa_m}{2}\right) \alpha_m + i\Lambda_m\,\frac{I(t)}{2e},\\
& I_\text{dc} = \frac{2eE_J}{\hbar} \,
\sin\left[ \frac{2eVt}{\hbar} - \sum_{m} \Lambda_{m} (\alpha_{m} + \alpha^*_{m}) \right].
\end{align}\end{subequations}
Their numerical solution by a fourth-order Runge-Kutta method yields average mode occupations $\bar{n}(\omega_m)=\overline{|\alpha_m(t)|^2}$ and the dc current $I_\text{dc}=\overline{I(t)}$,
with the time averaging denoted by the overbar.
}

The resulting $I-V$ curve, shown in Fig.~\ref{fig:IVcurves}, agrees with the results of the $P(E)$ theory in a large interval of voltages. 
\qqq{
At low voltages, both approaches are invalid; the rich low-voltage physics \cite{remez_bloch_2024} is beyond the scope of the present paper.
The classical calculation overestimates the width of the bistable region; indeed, it does not account for spontaneous emission of photons which tends to destabilize the cold state. On the other hand, the kinetic equation does not allow for switching between the two stable states. Estimation of the switching rate, relevant for experimental observability of the hysteresis, but requiring quite different approaches, is a separate problem that we prefer to relegate to a future study.
}

\qqq{The amplitudes $\alpha_m(t)$ have chaotic dynamics in the hot state (see Appendix~\ref{app:chaos}).}
Chaos is due to a nonlinear element (the small~JJ), and is absent if the system is treated in the linear approximation, as in Ref.~\cite{rastelli_tunable_2018, pekola_heat_2024}. A curvature of the mode dispersion leading to $\delta_m\neq\mathrm{const}$ is also important. Indeed, a superconducting resonator with $\omega_m=m\omega_1$ was found to exhibit coherent dynamics~\cite{cassidy_demonstration_2017, simon_theory_2018}.
Replacing the JJ chain by a superconducting transmission line with $\omega_m\approx(m-1/2)\omega_1$ in our calculations, we also find for some voltages a non-thermal regime resembling a laser with phase diffusion, combined with the thermal regime at other voltages \emph{in the same structure} (Appendix~\ref{app:transmission_line}). Competition between chaotic and coherent behavior, and conditions for appearance of these regimes %
are beyond the scope of the present paper, but deserve further investigation.

\qqq{
\textit{Experimental signatures.----}
Photonic overheating can be probed by implementing the small junction as a SQUID; then $E_J$ can be varied by an external magnetic field.
Upon increasing~$E_J$, the $I-V$ should change its shape, extending towards higher voltages (while for an equilibrium bath, the shape is fixed by the $E_J$-independent $P(E)$ function, with only a vertical scaling $I\propto{E}_J^2$). The current sharply drops at a specific voltage, which is not associated with any intrinsic energy scale of the system, but depends on~$E_J$.
Alternatively, one can measure photons escaping into the external circuit~\cite{fraudet_direct_2024}: 
for $2eV<\hbar\omega_\text{p}$, the spectrum of photons emitted into a cold environment is confined between $0<\hbar\omega<2eV$, while an overheated environment emits photons in the whole range $0<\omega<\omega_\text{p}$, as seen from Fig.~\ref{fig:temperatures} at $2eV=0.5\,\hbar\omega_\text{p}$.}

\textit{Conclusions.---}
We have shown that a long JJ chain, coupled to a small voltage-biased junction has some characteristics of a good thermal bath: its dynamics is chaotic, the state of each mode is thermal, although thermalization is incomplete, since different modes may have different temperatures.
Most importantly, the state of this bath can be significantly affected by a small junction.
Namely, the photonic degrees of freedom of the bath become strongly overheated, similar to Joule heating of electronic bath by a current flowing in an electrical conductor. It is thus important to check for this possibility in circuit quantum electrodynamics experiments, where the photonic bath state is usually taken for granted.

We thank J.~Estève for drawing our attention to this problem. We also acknowledge illuminating discussions with O.~Buisson, N.~Crescini, and W.~Guichard. The support of the superconducting quantum circuits team of Institut Néel is warmly acknowledged. This work was supported by the French National Research Agency (ANR) in the framework of the TRIANGLE project (ANR-20-CE47-0011) and by the European Research Council (ERC) under Grant Agreement No. 101001310 466 (SuperProtected).

\bibliography{PRLbib}

\begin{thebibliography}{48}%
\makeatletter
\providecommand \@ifxundefined [1]{%
 \@ifx{#1\undefined}
}%
\providecommand \@ifnum [1]{%
 \ifnum #1\expandafter \@firstoftwo
 \else \expandafter \@secondoftwo
 \fi
}%
\providecommand \@ifx [1]{%
 \ifx #1\expandafter \@firstoftwo
 \else \expandafter \@secondoftwo
 \fi
}%
\providecommand \natexlab [1]{#1}%
\providecommand \enquote  [1]{``#1''}%
\providecommand \bibnamefont  [1]{#1}%
\providecommand \bibfnamefont [1]{#1}%
\providecommand \citenamefont [1]{#1}%
\providecommand \href@noop [0]{\@secondoftwo}%
\providecommand \href [0]{\begingroup \@sanitize@url \@href}%
\providecommand \@href[1]{\@@startlink{#1}\@@href}%
\providecommand \@@href[1]{\endgroup#1\@@endlink}%
\providecommand \@sanitize@url [0]{\catcode `\\12\catcode `\$12\catcode
  `\&12\catcode `\#12\catcode `\^12\catcode `\_12\catcode `\%12\relax}%
\providecommand \@@startlink[1]{}%
\providecommand \@@endlink[0]{}%
\providecommand \url  [0]{\begingroup\@sanitize@url \@url }%
\providecommand \@url [1]{\endgroup\@href {#1}{\urlprefix }}%
\providecommand \urlprefix  [0]{URL }%
\providecommand \Eprint [0]{\href }%
\providecommand \doibase [0]{https://doi.org/}%
\providecommand \selectlanguage [0]{\@gobble}%
\providecommand \bibinfo  [0]{\@secondoftwo}%
\providecommand \bibfield  [0]{\@secondoftwo}%
\providecommand \translation [1]{[#1]}%
\providecommand \BibitemOpen [0]{}%
\providecommand \bibitemStop [0]{}%
\providecommand \bibitemNoStop [0]{.\EOS\space}%
\providecommand \EOS [0]{\spacefactor3000\relax}%
\providecommand \BibitemShut  [1]{\csname bibitem#1\endcsname}%
\let\auto@bib@innerbib\@empty
\bibitem [{\citenamefont {Harrington}\ \emph {et~al.}(2022)\citenamefont
  {Harrington}, \citenamefont {Mueller},\ and\ \citenamefont
  {Murch}}]{harrington_engineered_2022}%
  \BibitemOpen
  \bibfield  {author} {\bibinfo {author} {\bibfnamefont {P.~M.}\ \bibnamefont
  {Harrington}}, \bibinfo {author} {\bibfnamefont {E.}~\bibnamefont
  {Mueller}},\ and\ \bibinfo {author} {\bibfnamefont {K.}~\bibnamefont
  {Murch}},\ }\bibfield  {title} {\bibinfo {title} {Engineered {Dissipation}
  for {Quantum} {Information} {Science}},\ }\href
  {https://doi.org/10.1038/s42254-022-00494-8} {\bibfield  {journal} {\bibinfo
  {journal} {Nature Reviews Physics}\ }\textbf {\bibinfo {volume} {4}},\
  \bibinfo {pages} {660} (\bibinfo {year} {2022})}\BibitemShut {NoStop}%
\bibitem [{\citenamefont {Caldeira}\ and\ \citenamefont
  {Leggett}(1983)}]{caldeira_quantum_1983}%
  \BibitemOpen
  \bibfield  {author} {\bibinfo {author} {\bibfnamefont {A.}~\bibnamefont
  {Caldeira}}\ and\ \bibinfo {author} {\bibfnamefont {A.}~\bibnamefont
  {Leggett}},\ }\bibfield  {title} {\bibinfo {title} {Quantum tunnelling in a
  dissipative system},\ }\href
  {https://doi.org/https://doi.org/10.1016/0003-4916(83)90202-6} {\bibfield
  {journal} {\bibinfo  {journal} {Annals of Physics}\ }\textbf {\bibinfo
  {volume} {149}},\ \bibinfo {pages} {374} (\bibinfo {year}
  {1983})}\BibitemShut {NoStop}%
\bibitem [{\citenamefont {Gabelli}\ \emph {et~al.}(2004)\citenamefont
  {Gabelli}, \citenamefont {Reydellet}, \citenamefont {F\`eve}, \citenamefont
  {Berroir}, \citenamefont {Pla\ifmmode~\mbox{\c{c}}\else \c{c}\fi{}ais},
  \citenamefont {Roche},\ and\ \citenamefont {Glattli}}]{gabelli_hanbury_2004}%
  \BibitemOpen
  \bibfield  {author} {\bibinfo {author} {\bibfnamefont {J.}~\bibnamefont
  {Gabelli}}, \bibinfo {author} {\bibfnamefont {L.-H.}\ \bibnamefont
  {Reydellet}}, \bibinfo {author} {\bibfnamefont {G.}~\bibnamefont {F\`eve}},
  \bibinfo {author} {\bibfnamefont {J.-M.}\ \bibnamefont {Berroir}}, \bibinfo
  {author} {\bibfnamefont {B.}~\bibnamefont {Pla\ifmmode~\mbox{\c{c}}\else
  \c{c}\fi{}ais}}, \bibinfo {author} {\bibfnamefont {P.}~\bibnamefont
  {Roche}},\ and\ \bibinfo {author} {\bibfnamefont {D.~C.}\ \bibnamefont
  {Glattli}},\ }\bibfield  {title} {\bibinfo {title} {{Hanbury Brown--Twiss
  Correlations to Probe the Population Statistics of GHz Photons Emitted by
  Conductors}},\ }\href {https://doi.org/10.1103/PhysRevLett.93.056801}
  {\bibfield  {journal} {\bibinfo  {journal} {Phys. Rev. Lett.}\ }\textbf
  {\bibinfo {volume} {93}},\ \bibinfo {pages} {056801} (\bibinfo {year}
  {2004})}\BibitemShut {NoStop}%
\bibitem [{\citenamefont {Hofheinz}\ \emph {et~al.}(2011)\citenamefont
  {Hofheinz}, \citenamefont {Portier}, \citenamefont {Baudouin}, \citenamefont
  {Joyez}, \citenamefont {Vion}, \citenamefont {Bertet}, \citenamefont
  {Roche},\ and\ \citenamefont {Esteve}}]{hofheinz_bright_2011}%
  \BibitemOpen
  \bibfield  {author} {\bibinfo {author} {\bibfnamefont {M.}~\bibnamefont
  {Hofheinz}}, \bibinfo {author} {\bibfnamefont {F.}~\bibnamefont {Portier}},
  \bibinfo {author} {\bibfnamefont {Q.}~\bibnamefont {Baudouin}}, \bibinfo
  {author} {\bibfnamefont {P.}~\bibnamefont {Joyez}}, \bibinfo {author}
  {\bibfnamefont {D.}~\bibnamefont {Vion}}, \bibinfo {author} {\bibfnamefont
  {P.}~\bibnamefont {Bertet}}, \bibinfo {author} {\bibfnamefont
  {P.}~\bibnamefont {Roche}},\ and\ \bibinfo {author} {\bibfnamefont
  {D.}~\bibnamefont {Esteve}},\ }\bibfield  {title} {\bibinfo {title} {The
  {Bright} {Side} of {Coulomb} {Blockade}},\ }\href
  {https://doi.org/10.1103/PhysRevLett.106.217005} {\bibfield  {journal}
  {\bibinfo  {journal} {Phys. Rev. Lett.}\ }\textbf {\bibinfo {volume} {106}},\
  \bibinfo {pages} {217005} (\bibinfo {year} {2011})}\BibitemShut {NoStop}%
\bibitem [{\citenamefont {Fraudet}\ \emph {et~al.}(2024)\citenamefont
  {Fraudet}, \citenamefont {Snyman}, \citenamefont {Basko}, \citenamefont
  {L\'eger}, \citenamefont {S\'epulcre}, \citenamefont {Ranadive},
  \citenamefont {Le~Gal}, \citenamefont {Torras-Coloma}, \citenamefont
  {Florens},\ and\ \citenamefont {Roch}}]{fraudet_direct_2024}%
  \BibitemOpen
  \bibfield  {author} {\bibinfo {author} {\bibfnamefont {D.}~\bibnamefont
  {Fraudet}}, \bibinfo {author} {\bibfnamefont {I.}~\bibnamefont {Snyman}},
  \bibinfo {author} {\bibfnamefont {D.~M.}\ \bibnamefont {Basko}}, \bibinfo
  {author} {\bibfnamefont {S.}~\bibnamefont {L\'eger}}, \bibinfo {author}
  {\bibfnamefont {T.}~\bibnamefont {S\'epulcre}}, \bibinfo {author}
  {\bibfnamefont {A.}~\bibnamefont {Ranadive}}, \bibinfo {author}
  {\bibfnamefont {G.}~\bibnamefont {Le~Gal}}, \bibinfo {author} {\bibfnamefont
  {A.}~\bibnamefont {Torras-Coloma}}, \bibinfo {author} {\bibfnamefont
  {S.}~\bibnamefont {Florens}},\ and\ \bibinfo {author} {\bibfnamefont
  {N.}~\bibnamefont {Roch}},\ }\href@noop {} {\bibinfo {title} {Direct
  detection of down-converted photons spontaneously produced at a single
  {Josephson} junction}} (\bibinfo {year} {2024}),\ \Eprint
  {https://arxiv.org/abs/2405.00411} {arXiv:2405.00411 [cond-mat.mes-hall]}
  \BibitemShut {NoStop}%
\bibitem [{\citenamefont {Sundaresan}\ \emph {et~al.}(2015)\citenamefont
  {Sundaresan}, \citenamefont {Liu}, \citenamefont {Sadri}, \citenamefont
  {Szocs}, \citenamefont {Underwood}, \citenamefont {Malekakhlagh},
  \citenamefont {T{\"u}reci},\ and\ \citenamefont
  {Houck}}]{sundaresan_beyond_2015}%
  \BibitemOpen
  \bibfield  {author} {\bibinfo {author} {\bibfnamefont {N.~M.}\ \bibnamefont
  {Sundaresan}}, \bibinfo {author} {\bibfnamefont {Y.}~\bibnamefont {Liu}},
  \bibinfo {author} {\bibfnamefont {D.}~\bibnamefont {Sadri}}, \bibinfo
  {author} {\bibfnamefont {L.~J.}\ \bibnamefont {Szocs}}, \bibinfo {author}
  {\bibfnamefont {D.~L.}\ \bibnamefont {Underwood}}, \bibinfo {author}
  {\bibfnamefont {M.}~\bibnamefont {Malekakhlagh}}, \bibinfo {author}
  {\bibfnamefont {H.~E.}\ \bibnamefont {T{\"u}reci}},\ and\ \bibinfo {author}
  {\bibfnamefont {A.~A.}\ \bibnamefont {Houck}},\ }\bibfield  {title} {\bibinfo
  {title} {{Beyond Strong Coupling in a Multimode Cavity}},\ }\href
  {https://doi.org/10.1103/physrevx.5.021035} {\bibfield  {journal} {\bibinfo
  {journal} {Phys. Rev. X}\ }\textbf {\bibinfo {volume} {5}},\ \bibinfo {pages}
  {021035 } (\bibinfo {year} {2015})}\BibitemShut {NoStop}%
\bibitem [{\citenamefont {Jaako}\ \emph {et~al.}(2016)\citenamefont {Jaako},
  \citenamefont {Xiang}, \citenamefont {Garcia-Ripoll},\ and\ \citenamefont
  {Rabl}}]{Jaako2016}%
  \BibitemOpen
  \bibfield  {author} {\bibinfo {author} {\bibfnamefont {T.}~\bibnamefont
  {Jaako}}, \bibinfo {author} {\bibfnamefont {Z.-L.}\ \bibnamefont {Xiang}},
  \bibinfo {author} {\bibfnamefont {J.~J.}\ \bibnamefont {Garcia-Ripoll}},\
  and\ \bibinfo {author} {\bibfnamefont {P.}~\bibnamefont {Rabl}},\ }\bibfield
  {title} {\bibinfo {title} {{Ultrastrong-coupling phenomena beyond the Dicke
  model}},\ }\href {https://doi.org/10.1103/PhysRevA.94.033850} {\bibfield
  {journal} {\bibinfo  {journal} {Phys. Rev. A}\ }\textbf {\bibinfo {volume}
  {94}},\ \bibinfo {pages} {033850} (\bibinfo {year} {2016})}\BibitemShut
  {NoStop}%
\bibitem [{\citenamefont {Forn-Diaz}\ \emph {et~al.}(2017)\citenamefont
  {Forn-Diaz}, \citenamefont {Garcia-Ripoll}, \citenamefont {Peropadre},
  \citenamefont {Orgiazzi}, \citenamefont {Yurtalan}, \citenamefont
  {Belyansky}, \citenamefont {Wilson},\ and\ \citenamefont
  {Lupascu}}]{forndiaz_ultrastrong_2017}%
  \BibitemOpen
  \bibfield  {author} {\bibinfo {author} {\bibfnamefont {P.}~\bibnamefont
  {Forn-Diaz}}, \bibinfo {author} {\bibfnamefont {J.~J.}\ \bibnamefont
  {Garcia-Ripoll}}, \bibinfo {author} {\bibfnamefont {B.}~\bibnamefont
  {Peropadre}}, \bibinfo {author} {\bibfnamefont {J.-L.}\ \bibnamefont
  {Orgiazzi}}, \bibinfo {author} {\bibfnamefont {M.~A.}\ \bibnamefont
  {Yurtalan}}, \bibinfo {author} {\bibfnamefont {R.}~\bibnamefont {Belyansky}},
  \bibinfo {author} {\bibfnamefont {C.~M.}\ \bibnamefont {Wilson}},\ and\
  \bibinfo {author} {\bibfnamefont {A.}~\bibnamefont {Lupascu}},\ }\bibfield
  {title} {\bibinfo {title} {{Ultrastrong coupling of a single artificial atom
  to an electromagnetic continuum in the nonperturbative regime}},\ }\href
  {https://doi.org/10.1038/nphys3905} {\bibfield  {journal} {\bibinfo
  {journal} {Nature Physics}\ }\textbf {\bibinfo {volume} {13}},\ \bibinfo
  {pages} {39} (\bibinfo {year} {2017})}\BibitemShut {NoStop}%
\bibitem [{\citenamefont {Puertas~Martinez}\ \emph {et~al.}(2019)\citenamefont
  {Puertas~Martinez}, \citenamefont {Leger}, \citenamefont {Gheeraert},
  \citenamefont {Dassonneville}, \citenamefont {Planat}, \citenamefont
  {Foroughi}, \citenamefont {Krupko}, \citenamefont {Buisson}, \citenamefont
  {Naud}, \citenamefont {Hasch-Guichard}, \citenamefont {Florens},
  \citenamefont {Snyman},\ and\ \citenamefont
  {Roch}}]{puertasmartinez_tunable_2019}%
  \BibitemOpen
  \bibfield  {author} {\bibinfo {author} {\bibfnamefont {J.}~\bibnamefont
  {Puertas~Martinez}}, \bibinfo {author} {\bibfnamefont {S.}~\bibnamefont
  {Leger}}, \bibinfo {author} {\bibfnamefont {N.}~\bibnamefont {Gheeraert}},
  \bibinfo {author} {\bibfnamefont {R.}~\bibnamefont {Dassonneville}}, \bibinfo
  {author} {\bibfnamefont {L.}~\bibnamefont {Planat}}, \bibinfo {author}
  {\bibfnamefont {F.}~\bibnamefont {Foroughi}}, \bibinfo {author}
  {\bibfnamefont {Y.}~\bibnamefont {Krupko}}, \bibinfo {author} {\bibfnamefont
  {O.}~\bibnamefont {Buisson}}, \bibinfo {author} {\bibfnamefont
  {C.}~\bibnamefont {Naud}}, \bibinfo {author} {\bibfnamefont {W.}~\bibnamefont
  {Hasch-Guichard}}, \bibinfo {author} {\bibfnamefont {S.}~\bibnamefont
  {Florens}}, \bibinfo {author} {\bibfnamefont {I.}~\bibnamefont {Snyman}},\
  and\ \bibinfo {author} {\bibfnamefont {N.}~\bibnamefont {Roch}},\ }\bibfield
  {title} {\bibinfo {title} {{A tunable Josephson platform to explore many-body
  quantum optics in circuit-QED}},\ }\href
  {https://doi.org/10.1038/s41534-018-0104-0} {\bibfield  {journal} {\bibinfo
  {journal} {npj Quantum Information}\ }\textbf {\bibinfo {volume} {5}},\
  \bibinfo {pages} {1829} (\bibinfo {year} {2019})}\BibitemShut {NoStop}%
\bibitem [{\citenamefont {Kuzmin}\ \emph {et~al.}(2019)\citenamefont {Kuzmin},
  \citenamefont {Mehta}, \citenamefont {Grabon}, \citenamefont {Mencia},\ and\
  \citenamefont {Manucharyan}}]{kuzmin_superstrong_2019}%
  \BibitemOpen
  \bibfield  {author} {\bibinfo {author} {\bibfnamefont {R.}~\bibnamefont
  {Kuzmin}}, \bibinfo {author} {\bibfnamefont {N.}~\bibnamefont {Mehta}},
  \bibinfo {author} {\bibfnamefont {N.}~\bibnamefont {Grabon}}, \bibinfo
  {author} {\bibfnamefont {R.}~\bibnamefont {Mencia}},\ and\ \bibinfo {author}
  {\bibfnamefont {V.~E.}\ \bibnamefont {Manucharyan}},\ }\bibfield  {title}
  {\bibinfo {title} {{Superstrong coupling in circuit quantum
  electrodynamics}},\ }\href {https://doi.org/10.1038/s41534-019-0134-2}
  {\bibfield  {journal} {\bibinfo  {journal} {npj Quantum Information}\
  }\textbf {\bibinfo {volume} {5}},\ \bibinfo {pages} {1 } (\bibinfo {year}
  {2019})}\BibitemShut {NoStop}%
\bibitem [{\citenamefont {Kuzmin}\ \emph {et~al.}(2021)\citenamefont {Kuzmin},
  \citenamefont {Grabon}, \citenamefont {Mehta}, \citenamefont {Burshtein},
  \citenamefont {Goldstein}, \citenamefont {Houzet}, \citenamefont {Glazman},\
  and\ \citenamefont {Manucharyan}}]{kuzmin_inelastic_2021}%
  \BibitemOpen
  \bibfield  {author} {\bibinfo {author} {\bibfnamefont {R.}~\bibnamefont
  {Kuzmin}}, \bibinfo {author} {\bibfnamefont {N.}~\bibnamefont {Grabon}},
  \bibinfo {author} {\bibfnamefont {N.}~\bibnamefont {Mehta}}, \bibinfo
  {author} {\bibfnamefont {A.}~\bibnamefont {Burshtein}}, \bibinfo {author}
  {\bibfnamefont {M.}~\bibnamefont {Goldstein}}, \bibinfo {author}
  {\bibfnamefont {M.}~\bibnamefont {Houzet}}, \bibinfo {author} {\bibfnamefont
  {L.~I.}\ \bibnamefont {Glazman}},\ and\ \bibinfo {author} {\bibfnamefont
  {V.~E.}\ \bibnamefont {Manucharyan}},\ }\bibfield  {title} {\bibinfo {title}
  {{Inelastic Scattering of a Photon by a Quantum Phase Slip}},\ }\href
  {https://doi.org/10.1103/physrevlett.126.197701} {\bibfield  {journal}
  {\bibinfo  {journal} {Phys. Rev. Lett.}\ }\textbf {\bibinfo {volume} {126}},\
  \bibinfo {pages} {197701} (\bibinfo {year} {2021})}\BibitemShut {NoStop}%
\bibitem [{\citenamefont {L\'eger}\ \emph {et~al.}(2023)\citenamefont
  {L\'eger}, \citenamefont {S\'epulcre}, \citenamefont {Fraudet}, \citenamefont
  {Buisson}, \citenamefont {Naud}, \citenamefont {Hasch-Guichard},
  \citenamefont {Florens}, \citenamefont {Snyman}, \citenamefont {Basko},\ and\
  \citenamefont {Roch}}]{leger_revealing_2023}%
  \BibitemOpen
  \bibfield  {author} {\bibinfo {author} {\bibfnamefont {S.}~\bibnamefont
  {L\'eger}}, \bibinfo {author} {\bibfnamefont {T.}~\bibnamefont {S\'epulcre}},
  \bibinfo {author} {\bibfnamefont {D.}~\bibnamefont {Fraudet}}, \bibinfo
  {author} {\bibfnamefont {O.}~\bibnamefont {Buisson}}, \bibinfo {author}
  {\bibfnamefont {C.}~\bibnamefont {Naud}}, \bibinfo {author} {\bibfnamefont
  {W.}~\bibnamefont {Hasch-Guichard}}, \bibinfo {author} {\bibfnamefont
  {S.}~\bibnamefont {Florens}}, \bibinfo {author} {\bibfnamefont
  {I.}~\bibnamefont {Snyman}}, \bibinfo {author} {\bibfnamefont {D.~M.}\
  \bibnamefont {Basko}},\ and\ \bibinfo {author} {\bibfnamefont
  {N.}~\bibnamefont {Roch}},\ }\bibfield  {title} {\bibinfo {title} {{Revealing
  the finite-frequency response of a bosonic quantum impurity}},\ }\href
  {https://doi.org/10.21468/SciPostPhys.14.5.130} {\bibfield  {journal}
  {\bibinfo  {journal} {SciPost Phys.}\ }\textbf {\bibinfo {volume} {14}},\
  \bibinfo {pages} {130} (\bibinfo {year} {2023})}\BibitemShut {NoStop}%
\bibitem [{\citenamefont {Crescini}\ \emph {et~al.}(2023)\citenamefont
  {Crescini}, \citenamefont {Cailleaux}, \citenamefont {Guichard},
  \citenamefont {Naud}, \citenamefont {Buisson}, \citenamefont {W.~Murch},\
  and\ \citenamefont {Roch}}]{Crescini2023}%
  \BibitemOpen
  \bibfield  {author} {\bibinfo {author} {\bibfnamefont {N.}~\bibnamefont
  {Crescini}}, \bibinfo {author} {\bibfnamefont {S.}~\bibnamefont {Cailleaux}},
  \bibinfo {author} {\bibfnamefont {W.}~\bibnamefont {Guichard}}, \bibinfo
  {author} {\bibfnamefont {C.}~\bibnamefont {Naud}}, \bibinfo {author}
  {\bibfnamefont {O.}~\bibnamefont {Buisson}}, \bibinfo {author} {\bibfnamefont
  {K.}~\bibnamefont {W.~Murch}},\ and\ \bibinfo {author} {\bibfnamefont
  {N.}~\bibnamefont {Roch}},\ }\bibfield  {title} {\bibinfo {title} {{Evidence
  of dual Shapiro steps in a Josephson junction array}},\ }\href
  {https://doi.org/10.1038/s41567-023-01961-4} {\bibfield  {journal} {\bibinfo
  {journal} {Nature Physics}\ }\textbf {\bibinfo {volume} {19}},\ \bibinfo
  {pages} {851} (\bibinfo {year} {2023})}\BibitemShut {NoStop}%
\bibitem [{\citenamefont {Corlevi}\ \emph {et~al.}(2006)\citenamefont
  {Corlevi}, \citenamefont {Guichard}, \citenamefont {Hekking},\ and\
  \citenamefont {Haviland}}]{corlevi2006phase}%
  \BibitemOpen
  \bibfield  {author} {\bibinfo {author} {\bibfnamefont {S.}~\bibnamefont
  {Corlevi}}, \bibinfo {author} {\bibfnamefont {W.}~\bibnamefont {Guichard}},
  \bibinfo {author} {\bibfnamefont {F.~W.}\ \bibnamefont {Hekking}},\ and\
  \bibinfo {author} {\bibfnamefont {D.~B.}\ \bibnamefont {Haviland}},\
  }\bibfield  {title} {\bibinfo {title} {{Phase-charge duality of a Josephson
  junction in a fluctuating electromagnetic environment}},\ }\href
  {https://doi.org/10.1103/PhysRevLett.97.096802} {\bibfield  {journal}
  {\bibinfo  {journal} {Phys. Rev. Lett.}\ }\textbf {\bibinfo {volume} {97}},\
  \bibinfo {pages} {096802} (\bibinfo {year} {2006})}\BibitemShut {NoStop}%
\bibitem [{\citenamefont {Giacomelli}\ and\ \citenamefont
  {Ciuti}(2024)}]{giacomelli2024emergent}%
  \BibitemOpen
  \bibfield  {author} {\bibinfo {author} {\bibfnamefont {L.}~\bibnamefont
  {Giacomelli}}\ and\ \bibinfo {author} {\bibfnamefont {C.}~\bibnamefont
  {Ciuti}},\ }\bibfield  {title} {\bibinfo {title} {{Emergent quantum phase
  transition of a Josephson junction coupled to a high-impedance multimode
  resonator}},\ }\href {https://doi.org/10.1038/s41467-024-48558-w} {\bibfield
  {journal} {\bibinfo  {journal} {Nature Communications}\ }\textbf {\bibinfo
  {volume} {15}},\ \bibinfo {pages} {5455} (\bibinfo {year}
  {2024})}\BibitemShut {NoStop}%
\bibitem [{\citenamefont {Castellanos-Beltran}\ \emph
  {et~al.}(2008)\citenamefont {Castellanos-Beltran}, \citenamefont {Irwin},
  \citenamefont {Hilton}, \citenamefont {Vale},\ and\ \citenamefont
  {Lehnert}}]{Castellanos-Beltran2008}%
  \BibitemOpen
  \bibfield  {author} {\bibinfo {author} {\bibfnamefont {M.~A.}\ \bibnamefont
  {Castellanos-Beltran}}, \bibinfo {author} {\bibfnamefont {K.~D.}\
  \bibnamefont {Irwin}}, \bibinfo {author} {\bibfnamefont {G.~C.}\ \bibnamefont
  {Hilton}}, \bibinfo {author} {\bibfnamefont {L.~R.}\ \bibnamefont {Vale}},\
  and\ \bibinfo {author} {\bibfnamefont {K.~W.}\ \bibnamefont {Lehnert}},\
  }\bibfield  {title} {\bibinfo {title} {{Amplification and squeezing of
  quantum noise with a tunable Josephson metamaterial}},\ }\href
  {https://doi.org/10.1038/nphys1090} {\bibfield  {journal} {\bibinfo
  {journal} {Nature Physics}\ }\textbf {\bibinfo {volume} {4}},\ \bibinfo
  {pages} {929} (\bibinfo {year} {2008})}\BibitemShut {NoStop}%
\bibitem [{\citenamefont {Esposito}\ \emph {et~al.}(2021)\citenamefont
  {Esposito}, \citenamefont {Ranadive}, \citenamefont {Planat},\ and\
  \citenamefont {Roch}}]{Esposito2021}%
  \BibitemOpen
  \bibfield  {author} {\bibinfo {author} {\bibfnamefont {M.}~\bibnamefont
  {Esposito}}, \bibinfo {author} {\bibfnamefont {A.}~\bibnamefont {Ranadive}},
  \bibinfo {author} {\bibfnamefont {L.}~\bibnamefont {Planat}},\ and\ \bibinfo
  {author} {\bibfnamefont {N.}~\bibnamefont {Roch}},\ }\bibfield  {title}
  {\bibinfo {title} {{Perspective on traveling wave microwave parametric
  amplifiers}},\ }\href {https://doi.org/10.1063/5.0064892} {\bibfield
  {journal} {\bibinfo  {journal} {Applied Physics Letters}\ }\textbf {\bibinfo
  {volume} {119}},\ \bibinfo {pages} {120501} (\bibinfo {year}
  {2021})}\BibitemShut {NoStop}%
\bibitem [{\citenamefont
  {Kitaev}(2006)}]{kitaev2006protectedqubitbasedsuperconducting}%
  \BibitemOpen
  \bibfield  {author} {\bibinfo {author} {\bibfnamefont {A.}~\bibnamefont
  {Kitaev}},\ }\href@noop {} {\bibinfo {title} {Protected qubit based on a
  superconducting current mirror}} (\bibinfo {year} {2006}),\ \Eprint
  {https://arxiv.org/abs/cond-mat/0609441} {arXiv:cond-mat/0609441
  [cond-mat.mes-hall]} \BibitemShut {NoStop}%
\bibitem [{\citenamefont {Nataf}\ and\ \citenamefont
  {Ciuti}(2011)}]{Nataf2011}%
  \BibitemOpen
  \bibfield  {author} {\bibinfo {author} {\bibfnamefont {P.}~\bibnamefont
  {Nataf}}\ and\ \bibinfo {author} {\bibfnamefont {C.}~\bibnamefont {Ciuti}},\
  }\bibfield  {title} {\bibinfo {title} {{Protected Quantum Computation with
  Multiple Resonators in Ultrastrong Coupling Circuit QED}},\ }\href
  {https://doi.org/10.1103/PhysRevLett.107.190402} {\bibfield  {journal}
  {\bibinfo  {journal} {Phys. Rev. Lett.}\ }\textbf {\bibinfo {volume} {107}},\
  \bibinfo {pages} {190402} (\bibinfo {year} {2011})}\BibitemShut {NoStop}%
\bibitem [{\citenamefont {Liu}\ and\ \citenamefont {Houck}(2017)}]{Liu2017}%
  \BibitemOpen
  \bibfield  {author} {\bibinfo {author} {\bibfnamefont {Y.}~\bibnamefont
  {Liu}}\ and\ \bibinfo {author} {\bibfnamefont {A.~A.}\ \bibnamefont
  {Houck}},\ }\bibfield  {title} {\bibinfo {title} {Quantum electrodynamics
  near a photonic bandgap},\ }\href {https://doi.org/10.1038/nphys3834}
  {\bibfield  {journal} {\bibinfo  {journal} {Nature Physics}\ }\textbf
  {\bibinfo {volume} {13}},\ \bibinfo {pages} {48} (\bibinfo {year}
  {2017})}\BibitemShut {NoStop}%
\bibitem [{\citenamefont {Mirhosseini}\ \emph {et~al.}(2018)\citenamefont
  {Mirhosseini}, \citenamefont {Kim}, \citenamefont {Ferreira}, \citenamefont
  {Kalaee}, \citenamefont {Sipahigil}, \citenamefont {Keller},\ and\
  \citenamefont {Painter}}]{Mirhosseini2018}%
  \BibitemOpen
  \bibfield  {author} {\bibinfo {author} {\bibfnamefont {M.}~\bibnamefont
  {Mirhosseini}}, \bibinfo {author} {\bibfnamefont {E.}~\bibnamefont {Kim}},
  \bibinfo {author} {\bibfnamefont {V.~S.}\ \bibnamefont {Ferreira}}, \bibinfo
  {author} {\bibfnamefont {M.}~\bibnamefont {Kalaee}}, \bibinfo {author}
  {\bibfnamefont {A.}~\bibnamefont {Sipahigil}}, \bibinfo {author}
  {\bibfnamefont {A.~J.}\ \bibnamefont {Keller}},\ and\ \bibinfo {author}
  {\bibfnamefont {O.}~\bibnamefont {Painter}},\ }\bibfield  {title} {\bibinfo
  {title} {Superconducting metamaterials for waveguide quantum
  electrodynamics},\ }\href {https://doi.org/10.1038/s41467-018-06142-z}
  {\bibfield  {journal} {\bibinfo  {journal} {Nature Communications}\ }\textbf
  {\bibinfo {volume} {9}},\ \bibinfo {pages} {3706} (\bibinfo {year}
  {2018})}\BibitemShut {NoStop}%
\bibitem [{\citenamefont {Sinha}\ \emph {et~al.}(2022)\citenamefont {Sinha},
  \citenamefont {Khan}, \citenamefont {C\"uce},\ and\ \citenamefont
  {T\"ureci}}]{Sinha2022}%
  \BibitemOpen
  \bibfield  {author} {\bibinfo {author} {\bibfnamefont {K.}~\bibnamefont
  {Sinha}}, \bibinfo {author} {\bibfnamefont {S.~A.}\ \bibnamefont {Khan}},
  \bibinfo {author} {\bibfnamefont {E.}~\bibnamefont {C\"uce}},\ and\ \bibinfo
  {author} {\bibfnamefont {H.~E.}\ \bibnamefont {T\"ureci}},\ }\bibfield
  {title} {\bibinfo {title} {{Radiative properties of an artificial atom
  coupled to a Josephson-junction array}},\ }\href
  {https://doi.org/10.1103/PhysRevA.106.033714} {\bibfield  {journal} {\bibinfo
   {journal} {Phys. Rev. A}\ }\textbf {\bibinfo {volume} {106}},\ \bibinfo
  {pages} {033714} (\bibinfo {year} {2022})}\BibitemShut {NoStop}%
\bibitem [{\citenamefont {Zhang}\ \emph {et~al.}(2023)\citenamefont {Zhang},
  \citenamefont {Kim}, \citenamefont {Mark}, \citenamefont {Choi},\ and\
  \citenamefont {Painter}}]{Zhang2023}%
  \BibitemOpen
  \bibfield  {author} {\bibinfo {author} {\bibfnamefont {X.}~\bibnamefont
  {Zhang}}, \bibinfo {author} {\bibfnamefont {E.}~\bibnamefont {Kim}}, \bibinfo
  {author} {\bibfnamefont {D.~K.}\ \bibnamefont {Mark}}, \bibinfo {author}
  {\bibfnamefont {S.}~\bibnamefont {Choi}},\ and\ \bibinfo {author}
  {\bibfnamefont {O.}~\bibnamefont {Painter}},\ }\bibfield  {title} {\bibinfo
  {title} {A superconducting quantum simulator based on a photonic-bandgap
  metamaterial},\ }\href {https://doi.org/10.1126/science.ade7651} {\bibfield
  {journal} {\bibinfo  {journal} {Science}\ }\textbf {\bibinfo {volume}
  {379}},\ \bibinfo {pages} {278} (\bibinfo {year} {2023})}\BibitemShut
  {NoStop}%
\bibitem [{\citenamefont {Ingold}\ and\ \citenamefont
  {Nazarov}(1992)}]{ingold_charge_2005}%
  \BibitemOpen
  \bibfield  {author} {\bibinfo {author} {\bibfnamefont {G.-L.}\ \bibnamefont
  {Ingold}}\ and\ \bibinfo {author} {\bibfnamefont {Y.~V.}\ \bibnamefont
  {Nazarov}},\ }\bibfield  {title} {\bibinfo {title} {Charge {Tunneling}
  {Rates} in {Ultrasmall} {Junctions}},\ }in\ \href
  {http://arxiv.org/abs/cond-mat/0508728} {\emph {\bibinfo {booktitle} {Single
  {Charge} {Tunneling}}}},\ \bibinfo {series} {NATO ASI Series B}, Vol.\
  \bibinfo {volume} {294},\ \bibinfo {editor} {edited by\ \bibinfo {editor}
  {\bibfnamefont {H.}~\bibnamefont {Grabert}}\ and\ \bibinfo {editor}
  {\bibfnamefont {M.~H.}\ \bibnamefont {Devoret}}}\ (\bibinfo  {publisher}
  {Plenum Press, New York},\ \bibinfo {year} {1992})\ pp.\ \bibinfo {pages}
  {21--107}\BibitemShut {NoStop}%
\bibitem [{\citenamefont {Gurevich}\ and\ \citenamefont
  {Mints}(1987)}]{gurevich_selfheating_1987}%
  \BibitemOpen
  \bibfield  {author} {\bibinfo {author} {\bibfnamefont {A.~V.}\ \bibnamefont
  {Gurevich}}\ and\ \bibinfo {author} {\bibfnamefont {R.~G.}\ \bibnamefont
  {Mints}},\ }\bibfield  {title} {\bibinfo {title} {Self-heating in normal
  metals and superconductors},\ }\href
  {https://doi.org/10.1103/RevModPhys.59.941} {\bibfield  {journal} {\bibinfo
  {journal} {Rev. Mod. Phys.}\ }\textbf {\bibinfo {volume} {59}},\ \bibinfo
  {pages} {941} (\bibinfo {year} {1987})}\BibitemShut {NoStop}%
\bibitem [{\citenamefont {Masluk}\ \emph {et~al.}(2012)\citenamefont {Masluk},
  \citenamefont {Pop}, \citenamefont {Kamal}, \citenamefont {Minev},\ and\
  \citenamefont {Devoret}}]{masluk_microwave_2012}%
  \BibitemOpen
  \bibfield  {author} {\bibinfo {author} {\bibfnamefont {N.~A.}\ \bibnamefont
  {Masluk}}, \bibinfo {author} {\bibfnamefont {I.~M.}\ \bibnamefont {Pop}},
  \bibinfo {author} {\bibfnamefont {A.}~\bibnamefont {Kamal}}, \bibinfo
  {author} {\bibfnamefont {Z.~K.}\ \bibnamefont {Minev}},\ and\ \bibinfo
  {author} {\bibfnamefont {M.~H.}\ \bibnamefont {Devoret}},\ }\bibfield
  {title} {\bibinfo {title} {{Microwave Characterization of Josephson Junction
  Arrays: Implementing a Low Loss Superinductance}},\ }\href
  {https://doi.org/10.1103/PhysRevLett.109.137002} {\bibfield  {journal}
  {\bibinfo  {journal} {Phys. Rev. Lett.}\ }\textbf {\bibinfo {volume} {109}},\
  \bibinfo {pages} {137002} (\bibinfo {year} {2012})}\BibitemShut {NoStop}%
\bibitem [{\citenamefont {Armour}\ \emph {et~al.}(2013)\citenamefont {Armour},
  \citenamefont {Blencowe}, \citenamefont {Brahimi},\ and\ \citenamefont
  {Rimberg}}]{armour_universal_2013}%
  \BibitemOpen
  \bibfield  {author} {\bibinfo {author} {\bibfnamefont {A.~D.}\ \bibnamefont
  {Armour}}, \bibinfo {author} {\bibfnamefont {M.~P.}\ \bibnamefont
  {Blencowe}}, \bibinfo {author} {\bibfnamefont {E.}~\bibnamefont {Brahimi}},\
  and\ \bibinfo {author} {\bibfnamefont {A.~J.}\ \bibnamefont {Rimberg}},\
  }\bibfield  {title} {\bibinfo {title} {{Universal Quantum Fluctuations of a
  Cavity Mode Driven by a Josephson Junction}},\ }\href
  {https://doi.org/10.1103/PhysRevLett.111.247001} {\bibfield  {journal}
  {\bibinfo  {journal} {Phys. Rev. Lett.}\ }\textbf {\bibinfo {volume} {111}},\
  \bibinfo {pages} {247001} (\bibinfo {year} {2013})}\BibitemShut {NoStop}%
\bibitem [{\citenamefont {Armour}\ \emph {et~al.}(2015)\citenamefont {Armour},
  \citenamefont {Kubala},\ and\ \citenamefont
  {Ankerhold}}]{armour_josephson_2015}%
  \BibitemOpen
  \bibfield  {author} {\bibinfo {author} {\bibfnamefont {A.~D.}\ \bibnamefont
  {Armour}}, \bibinfo {author} {\bibfnamefont {B.}~\bibnamefont {Kubala}},\
  and\ \bibinfo {author} {\bibfnamefont {J.}~\bibnamefont {Ankerhold}},\
  }\bibfield  {title} {\bibinfo {title} {Josephson photonics with a two-mode
  superconducting circuit},\ }\href
  {https://doi.org/10.1103/PhysRevB.91.184508} {\bibfield  {journal} {\bibinfo
  {journal} {Phys. Rev. B}\ }\textbf {\bibinfo {volume} {91}},\ \bibinfo
  {pages} {184508} (\bibinfo {year} {2015})}\BibitemShut {NoStop}%
\bibitem [{\citenamefont {Trif}\ and\ \citenamefont
  {Simon}(2015)}]{trif_photon_2015}%
  \BibitemOpen
  \bibfield  {author} {\bibinfo {author} {\bibfnamefont {M.}~\bibnamefont
  {Trif}}\ and\ \bibinfo {author} {\bibfnamefont {P.}~\bibnamefont {Simon}},\
  }\bibfield  {title} {\bibinfo {title} {Photon cross-correlations emitted by a
  josephson junction in two microwave cavities},\ }\href
  {https://doi.org/10.1103/PhysRevB.92.014503} {\bibfield  {journal} {\bibinfo
  {journal} {Phys. Rev. B}\ }\textbf {\bibinfo {volume} {92}},\ \bibinfo
  {pages} {014503} (\bibinfo {year} {2015})}\BibitemShut {NoStop}%
\bibitem [{\citenamefont {Hofer}\ \emph {et~al.}(2016)\citenamefont {Hofer},
  \citenamefont {Souquet},\ and\ \citenamefont {Clerk}}]{hofer_quantum_2016}%
  \BibitemOpen
  \bibfield  {author} {\bibinfo {author} {\bibfnamefont {P.~P.}\ \bibnamefont
  {Hofer}}, \bibinfo {author} {\bibfnamefont {J.-R.}\ \bibnamefont {Souquet}},\
  and\ \bibinfo {author} {\bibfnamefont {A.~A.}\ \bibnamefont {Clerk}},\
  }\bibfield  {title} {\bibinfo {title} {Quantum heat engine based on
  photon-assisted cooper pair tunneling},\ }\href
  {https://doi.org/10.1103/PhysRevB.93.041418} {\bibfield  {journal} {\bibinfo
  {journal} {Phys. Rev. B}\ }\textbf {\bibinfo {volume} {93}},\ \bibinfo
  {pages} {041418} (\bibinfo {year} {2016})}\BibitemShut {NoStop}%
\bibitem [{\citenamefont {Dambach}\ \emph {et~al.}(2017)\citenamefont
  {Dambach}, \citenamefont {Kubala},\ and\ \citenamefont
  {Ankerhold}}]{dambach_generating_2017}%
  \BibitemOpen
  \bibfield  {author} {\bibinfo {author} {\bibfnamefont {S.}~\bibnamefont
  {Dambach}}, \bibinfo {author} {\bibfnamefont {B.}~\bibnamefont {Kubala}},\
  and\ \bibinfo {author} {\bibfnamefont {J.}~\bibnamefont {Ankerhold}},\
  }\bibfield  {title} {\bibinfo {title} {{Generating entangled quantum
  microwaves in a Josephson-photonics device}},\ }\href
  {https://doi.org/10.1088/1367-2630/aa5bb6} {\bibfield  {journal} {\bibinfo
  {journal} {New Journal of Physics}\ }\textbf {\bibinfo {volume} {19}},\
  \bibinfo {pages} {023027} (\bibinfo {year} {2017})}\BibitemShut {NoStop}%
\bibitem [{\citenamefont {Cassidy}\ \emph {et~al.}(2017)\citenamefont
  {Cassidy}, \citenamefont {Bruno}, \citenamefont {Rubbert}, \citenamefont
  {Irfan}, \citenamefont {Kammhuber}, \citenamefont {Schouten}, \citenamefont
  {Akhmerov},\ and\ \citenamefont {Kouwenhoven}}]{cassidy_demonstration_2017}%
  \BibitemOpen
  \bibfield  {author} {\bibinfo {author} {\bibfnamefont {M.~C.}\ \bibnamefont
  {Cassidy}}, \bibinfo {author} {\bibfnamefont {A.}~\bibnamefont {Bruno}},
  \bibinfo {author} {\bibfnamefont {S.}~\bibnamefont {Rubbert}}, \bibinfo
  {author} {\bibfnamefont {M.}~\bibnamefont {Irfan}}, \bibinfo {author}
  {\bibfnamefont {J.}~\bibnamefont {Kammhuber}}, \bibinfo {author}
  {\bibfnamefont {R.~N.}\ \bibnamefont {Schouten}}, \bibinfo {author}
  {\bibfnamefont {A.~R.}\ \bibnamefont {Akhmerov}},\ and\ \bibinfo {author}
  {\bibfnamefont {L.~P.}\ \bibnamefont {Kouwenhoven}},\ }\bibfield  {title}
  {\bibinfo {title} {Demonstration of an ac {Josephson} junction laser},\
  }\href {https://doi.org/10.1126/science.aah6640} {\bibfield  {journal}
  {\bibinfo  {journal} {Science}\ }\textbf {\bibinfo {volume} {355}},\ \bibinfo
  {pages} {939} (\bibinfo {year} {2017})}\BibitemShut {NoStop}%
\bibitem [{\citenamefont {Simon}\ and\ \citenamefont
  {Cooper}(2018)}]{simon_theory_2018}%
  \BibitemOpen
  \bibfield  {author} {\bibinfo {author} {\bibfnamefont {S.~H.}\ \bibnamefont
  {Simon}}\ and\ \bibinfo {author} {\bibfnamefont {N.~R.}\ \bibnamefont
  {Cooper}},\ }\bibfield  {title} {\bibinfo {title} {{Theory of the Josephson
  Junction Laser}},\ }\href {https://doi.org/10.1103/PhysRevLett.121.027004}
  {\bibfield  {journal} {\bibinfo  {journal} {Phys. Rev. Lett.}\ }\textbf
  {\bibinfo {volume} {121}},\ \bibinfo {pages} {027004} (\bibinfo {year}
  {2018})}\BibitemShut {NoStop}%
\bibitem [{\citenamefont {Aiello}(2020)}]{aiello2020thesis}%
  \BibitemOpen
  \bibfield  {author} {\bibinfo {author} {\bibfnamefont {G.}~\bibnamefont
  {Aiello}},\ }\emph {\bibinfo {title} {Quantum dynamics of a high impedance
  microwave cavity strongly coupled to a Josephson junction}},\ \href
  {https://theses.hal.science/tel-03165358} {\bibinfo {type} {Phd thesis}},\
  \bibinfo  {school} {{Universit{\'e} Paris-Saclay}} (\bibinfo {year}
  {2020})\BibitemShut {NoStop}%
\bibitem [{\citenamefont {Lang}\ \emph {et~al.}(2023)\citenamefont {Lang},
  \citenamefont {Morley},\ and\ \citenamefont {Armour}}]{lang_discrete_2023}%
  \BibitemOpen
  \bibfield  {author} {\bibinfo {author} {\bibfnamefont {B.}~\bibnamefont
  {Lang}}, \bibinfo {author} {\bibfnamefont {G.~F.}\ \bibnamefont {Morley}},\
  and\ \bibinfo {author} {\bibfnamefont {A.~D.}\ \bibnamefont {Armour}},\
  }\bibfield  {title} {\bibinfo {title} {{Discrete time translation symmetry
  breaking in a Josephson junction laser}},\ }\href
  {https://doi.org/10.1103/PhysRevB.107.144509} {\bibfield  {journal} {\bibinfo
   {journal} {Phys. Rev. B}\ }\textbf {\bibinfo {volume} {107}},\ \bibinfo
  {pages} {144509} (\bibinfo {year} {2023})}\BibitemShut {NoStop}%
\bibitem [{\citenamefont {Subero}\ \emph {et~al.}(2023)\citenamefont {Subero},
  \citenamefont {Maillet}, \citenamefont {Golubev}, \citenamefont {Thomas},
  \citenamefont {Peltonen}, \citenamefont {Karimi}, \citenamefont
  {Mar{\'i}n-Su{\'a}rez}, \citenamefont {Yeyati}, \citenamefont {S{\'a}nchez},
  \citenamefont {Park},\ and\ \citenamefont {Pekola}}]{subero_bolometric_2023}%
  \BibitemOpen
  \bibfield  {author} {\bibinfo {author} {\bibfnamefont {D.}~\bibnamefont
  {Subero}}, \bibinfo {author} {\bibfnamefont {O.}~\bibnamefont {Maillet}},
  \bibinfo {author} {\bibfnamefont {D.~S.}\ \bibnamefont {Golubev}}, \bibinfo
  {author} {\bibfnamefont {G.}~\bibnamefont {Thomas}}, \bibinfo {author}
  {\bibfnamefont {J.~T.}\ \bibnamefont {Peltonen}}, \bibinfo {author}
  {\bibfnamefont {B.}~\bibnamefont {Karimi}}, \bibinfo {author} {\bibfnamefont
  {M.}~\bibnamefont {Mar{\'i}n-Su{\'a}rez}}, \bibinfo {author} {\bibfnamefont
  {A.~L.}\ \bibnamefont {Yeyati}}, \bibinfo {author} {\bibfnamefont
  {R.}~\bibnamefont {S{\'a}nchez}}, \bibinfo {author} {\bibfnamefont
  {S.}~\bibnamefont {Park}},\ and\ \bibinfo {author} {\bibfnamefont {J.~P.}\
  \bibnamefont {Pekola}},\ }\bibfield  {title} {\bibinfo {title} {{Bolometric
  detection of Josephson inductance in a highly resistive environment}},\
  }\href {https://doi.org/10.1038/s41467-023-43668-3} {\bibfield  {journal}
  {\bibinfo  {journal} {Nature Communications}\ }\textbf {\bibinfo {volume}
  {14}},\ \bibinfo {pages} {7924} (\bibinfo {year} {2023})}\BibitemShut
  {NoStop}%
\bibitem [{\citenamefont {Catelani}\ and\ \citenamefont
  {Basko}(2019)}]{catelani_nonequilibrium_2019}%
  \BibitemOpen
  \bibfield  {author} {\bibinfo {author} {\bibfnamefont {G.}~\bibnamefont
  {Catelani}}\ and\ \bibinfo {author} {\bibfnamefont {D.~M.}\ \bibnamefont
  {Basko}},\ }\bibfield  {title} {\bibinfo {title} {{Non-equilibrium
  quasiparticles in superconducting circuits: photons vs. phonons}},\ }\href
  {https://doi.org/10.21468/SciPostPhys.6.1.013} {\bibfield  {journal}
  {\bibinfo  {journal} {SciPost Phys.}\ }\textbf {\bibinfo {volume} {6}},\
  \bibinfo {pages} {013} (\bibinfo {year} {2019})}\BibitemShut {NoStop}%
\bibitem [{\citenamefont {Remez}\ \emph {et~al.}(2024)\citenamefont {Remez},
  \citenamefont {Kurilovich}, \citenamefont {Rieger},\ and\ \citenamefont
  {Glazman}}]{remez_bloch_2024}%
  \BibitemOpen
  \bibfield  {author} {\bibinfo {author} {\bibfnamefont {B.}~\bibnamefont
  {Remez}}, \bibinfo {author} {\bibfnamefont {V.~D.}\ \bibnamefont
  {Kurilovich}}, \bibinfo {author} {\bibfnamefont {M.}~\bibnamefont {Rieger}},\
  and\ \bibinfo {author} {\bibfnamefont {L.~I.}\ \bibnamefont {Glazman}},\
  }\bibfield  {title} {\bibinfo {title} {Bloch oscillations in a transmon
  embedded in a resonant electromagnetic environment},\ }\href
  {https://doi.org/10.1103/PhysRevB.110.054508} {\bibfield  {journal} {\bibinfo
   {journal} {Phys. Rev. B}\ }\textbf {\bibinfo {volume} {110}},\ \bibinfo
  {pages} {054508} (\bibinfo {year} {2024})}\BibitemShut {NoStop}%
\bibitem [{\citenamefont {Rastelli}\ and\ \citenamefont
  {Pop}(2018)}]{rastelli_tunable_2018}%
  \BibitemOpen
  \bibfield  {author} {\bibinfo {author} {\bibfnamefont {G.}~\bibnamefont
  {Rastelli}}\ and\ \bibinfo {author} {\bibfnamefont {I.~M.}\ \bibnamefont
  {Pop}},\ }\bibfield  {title} {\bibinfo {title} {{Tunable ohmic environment
  using Josephson junction chains}},\ }\href
  {https://doi.org/10.1103/PhysRevB.97.205429} {\bibfield  {journal} {\bibinfo
  {journal} {Phys. Rev. B}\ }\textbf {\bibinfo {volume} {97}},\ \bibinfo
  {pages} {205429} (\bibinfo {year} {2018})}\BibitemShut {NoStop}%
\bibitem [{\citenamefont {Pekola}\ and\ \citenamefont
  {Karimi}(2024)}]{pekola_heat_2024}%
  \BibitemOpen
  \bibfield  {author} {\bibinfo {author} {\bibfnamefont {J.~P.}\ \bibnamefont
  {Pekola}}\ and\ \bibinfo {author} {\bibfnamefont {B.}~\bibnamefont
  {Karimi}},\ }\bibfield  {title} {\bibinfo {title} {{Heat Bath in a Quantum
  Circuit}},\ }\href {https://doi.org/10.3390/e26050429} {\bibfield  {journal}
  {\bibinfo  {journal} {Entropy}\ }\textbf {\bibinfo {volume} {26}},\ \bibinfo
  {pages} {429} (\bibinfo {year} {2024})}\BibitemShut {NoStop}%
\bibitem [{\citenamefont {Ismail}\ \emph {et~al.}(2016)\citenamefont {Ismail},
  \citenamefont {Kores}, \citenamefont {Geskus},\ and\ \citenamefont
  {Pollnau}}]{ismail_fabryperot_2016}%
  \BibitemOpen
  \bibfield  {author} {\bibinfo {author} {\bibfnamefont {N.}~\bibnamefont
  {Ismail}}, \bibinfo {author} {\bibfnamefont {C.~C.}\ \bibnamefont {Kores}},
  \bibinfo {author} {\bibfnamefont {D.}~\bibnamefont {Geskus}},\ and\ \bibinfo
  {author} {\bibfnamefont {M.}~\bibnamefont {Pollnau}},\ }\bibfield  {title}
  {\bibinfo {title} {{Fabry-Pérot resonator: spectral line shapes, generic and
  related Airy distributions, linewidths, finesses, and performance at low or
  frequency-dependent reflectivity}},\ }\href
  {https://doi.org/10.1364/OE.24.016366} {\bibfield  {journal} {\bibinfo
  {journal} {Opt. Express}\ }\textbf {\bibinfo {volume} {24}},\ \bibinfo
  {pages} {16366} (\bibinfo {year} {2016})}\BibitemShut {NoStop}%
\bibitem [{\citenamefont {Catelani}\ \emph
  {et~al.}(2011{\natexlab{a}})\citenamefont {Catelani}, \citenamefont {Koch},
  \citenamefont {Frunzio}, \citenamefont {Schoelkopf}, \citenamefont
  {Devoret},\ and\ \citenamefont {Glazman}}]{catelani_quasiparticle_2011}%
  \BibitemOpen
  \bibfield  {author} {\bibinfo {author} {\bibfnamefont {G.}~\bibnamefont
  {Catelani}}, \bibinfo {author} {\bibfnamefont {J.}~\bibnamefont {Koch}},
  \bibinfo {author} {\bibfnamefont {L.}~\bibnamefont {Frunzio}}, \bibinfo
  {author} {\bibfnamefont {R.~J.}\ \bibnamefont {Schoelkopf}}, \bibinfo
  {author} {\bibfnamefont {M.~H.}\ \bibnamefont {Devoret}},\ and\ \bibinfo
  {author} {\bibfnamefont {L.~I.}\ \bibnamefont {Glazman}},\ }\bibfield
  {title} {\bibinfo {title} {{Quasiparticle Relaxation of Superconducting
  Qubits in the Presence of Flux}},\ }\href
  {https://doi.org/10.1103/PhysRevLett.106.077002} {\bibfield  {journal}
  {\bibinfo  {journal} {Phys. Rev. Lett.}\ }\textbf {\bibinfo {volume} {106}},\
  \bibinfo {pages} {077002} (\bibinfo {year} {2011}{\natexlab{a}})}\BibitemShut
  {NoStop}%
\bibitem [{\citenamefont {Catelani}\ \emph
  {et~al.}(2011{\natexlab{b}})\citenamefont {Catelani}, \citenamefont
  {Schoelkopf}, \citenamefont {Devoret},\ and\ \citenamefont
  {Glazman}}]{catelani_relaxation_2011}%
  \BibitemOpen
  \bibfield  {author} {\bibinfo {author} {\bibfnamefont {G.}~\bibnamefont
  {Catelani}}, \bibinfo {author} {\bibfnamefont {R.~J.}\ \bibnamefont
  {Schoelkopf}}, \bibinfo {author} {\bibfnamefont {M.~H.}\ \bibnamefont
  {Devoret}},\ and\ \bibinfo {author} {\bibfnamefont {L.~I.}\ \bibnamefont
  {Glazman}},\ }\bibfield  {title} {\bibinfo {title} {Relaxation and frequency
  shifts induced by quasiparticles in superconducting qubits},\ }\href
  {https://doi.org/10.1103/PhysRevB.84.064517} {\bibfield  {journal} {\bibinfo
  {journal} {Phys. Rev. B}\ }\textbf {\bibinfo {volume} {84}},\ \bibinfo
  {pages} {064517} (\bibinfo {year} {2011}{\natexlab{b}})}\BibitemShut
  {NoStop}%
\bibitem [{\citenamefont {Ambegaokar}\ and\ \citenamefont
  {Baratoff}(1963)}]{ambegaokar_tunneling_1963}%
  \BibitemOpen
  \bibfield  {author} {\bibinfo {author} {\bibfnamefont {V.}~\bibnamefont
  {Ambegaokar}}\ and\ \bibinfo {author} {\bibfnamefont {A.}~\bibnamefont
  {Baratoff}},\ }\bibfield  {title} {\bibinfo {title} {Tunneling between
  superconductors},\ }\href {https://doi.org/10.1103/PhysRevLett.10.486}
  {\bibfield  {journal} {\bibinfo  {journal} {Phys. Rev. Lett.}\ }\textbf
  {\bibinfo {volume} {10}},\ \bibinfo {pages} {486} (\bibinfo {year}
  {1963})}\BibitemShut {NoStop}%
\bibitem [{\citenamefont {Moody}\ and\ \citenamefont
  {Paterson}(1981)}]{moody_qp_relaxation}%
  \BibitemOpen
  \bibfield  {author} {\bibinfo {author} {\bibfnamefont {M.~V.}\ \bibnamefont
  {Moody}}\ and\ \bibinfo {author} {\bibfnamefont {J.~L.}\ \bibnamefont
  {Paterson}},\ }\bibfield  {title} {\bibinfo {title} {{Quasiparticle
  relaxation times in clean Al films}},\ }\href
  {https://doi.org/10.1103/PhysRevB.23.133} {\bibfield  {journal} {\bibinfo
  {journal} {Phys. Rev. B}\ }\textbf {\bibinfo {volume} {23}},\ \bibinfo
  {pages} {133} (\bibinfo {year} {1981})}\BibitemShut {NoStop}%
\bibitem [{\citenamefont {Anthore}\ \emph {et~al.}(2003)\citenamefont
  {Anthore}, \citenamefont {Pothier},\ and\ \citenamefont
  {Esteve}}]{anthore_density_2003}%
  \BibitemOpen
  \bibfield  {author} {\bibinfo {author} {\bibfnamefont {A.}~\bibnamefont
  {Anthore}}, \bibinfo {author} {\bibfnamefont {H.}~\bibnamefont {Pothier}},\
  and\ \bibinfo {author} {\bibfnamefont {D.}~\bibnamefont {Esteve}},\
  }\bibfield  {title} {\bibinfo {title} {{Density of States in a Superconductor
  Carrying a Supercurrent}},\ }\href
  {https://doi.org/10.1103/PhysRevLett.90.127001} {\bibfield  {journal}
  {\bibinfo  {journal} {Phys. Rev. Lett.}\ }\textbf {\bibinfo {volume} {90}},\
  \bibinfo {pages} {127001} (\bibinfo {year} {2003})}\BibitemShut {NoStop}%
\bibitem [{\citenamefont {Basko}\ \emph {et~al.}(2006)\citenamefont {Basko},
  \citenamefont {Aleiner},\ and\ \citenamefont {Altshuler}}]{basko_metal_2006}%
  \BibitemOpen
  \bibfield  {author} {\bibinfo {author} {\bibfnamefont {D.}~\bibnamefont
  {Basko}}, \bibinfo {author} {\bibfnamefont {I.}~\bibnamefont {Aleiner}},\
  and\ \bibinfo {author} {\bibfnamefont {B.}~\bibnamefont {Altshuler}},\
  }\bibfield  {title} {\bibinfo {title} {Metal–insulator transition in a
  weakly interacting many-electron system with localized single-particle
  states},\ }\href {https://doi.org/https://doi.org/10.1016/j.aop.2005.11.014}
  {\bibfield  {journal} {\bibinfo  {journal} {Annals of Physics}\ }\textbf
  {\bibinfo {volume} {321}},\ \bibinfo {pages} {1126} (\bibinfo {year}
  {2006})}\BibitemShut {NoStop}%
\bibitem [{\citenamefont {Aiello}\ \emph {et~al.}(2022)\citenamefont {Aiello},
  \citenamefont {F{\'e}chant}, \citenamefont {Morvan}, \citenamefont {Basset},
  \citenamefont {Aprili}, \citenamefont {Gabelli},\ and\ \citenamefont
  {Est{\`e}ve}}]{aiello_quantum_2022}%
  \BibitemOpen
  \bibfield  {author} {\bibinfo {author} {\bibfnamefont {G.}~\bibnamefont
  {Aiello}}, \bibinfo {author} {\bibfnamefont {M.}~\bibnamefont {F{\'e}chant}},
  \bibinfo {author} {\bibfnamefont {A.}~\bibnamefont {Morvan}}, \bibinfo
  {author} {\bibfnamefont {J.}~\bibnamefont {Basset}}, \bibinfo {author}
  {\bibfnamefont {M.}~\bibnamefont {Aprili}}, \bibinfo {author} {\bibfnamefont
  {J.}~\bibnamefont {Gabelli}},\ and\ \bibinfo {author} {\bibfnamefont
  {J.}~\bibnamefont {Est{\`e}ve}},\ }\bibfield  {title} {\bibinfo {title}
  {Quantum bath engineering of a high impedance microwave mode through
  quasiparticle tunneling},\ }\href
  {https://doi.org/10.1038/s41467-022-34762-z} {\bibfield  {journal} {\bibinfo
  {journal} {Nature Communications}\ }\textbf {\bibinfo {volume} {13}},\
  \bibinfo {pages} {7146} (\bibinfo {year} {2022})}\BibitemShut {NoStop}%
\end{thebibliography}%

\appendix

\section{Frequencies and decay rates of the environment's modes}
\label{app:param_computation}

To obtain the resonant frequencies of a long chain of large Josephson junctions, we first note that the infinite chain is characterised by the bulk dispersion of the propagating photonic modes~\cite{masluk_microwave_2012},
\begin{equation}\label{eq:dispersion}
    \omega(k) = \frac{\omega_\text{p}k}{\sqrt{k^2+C_\text{g}/C}},
    \quad
    k(\omega)=\sqrt{\frac{C_\text{g}}C}\,\frac\omega{\sqrt{\omega_\text{p}^2 - \omega^2}},
\end{equation}
where $C$ and $C_\text{g}$ are the junction and the island capacitances of the chain, respectively, $\omega_\text{p}$ is the plasma frequency, and we assume the dimensionless wave vector $k\ll1$. 
The impedance of the infinite chain $Z(\omega)=Z_0(1-\omega^2/\omega_\text{p}^2)^{-1/2}$.

On one end the chain is connected to the external circuit with the impedance ${Z}_\text{ext}=50\:\Omega$, so the voltage amplitude reflection coefficient on this end is given by $[Z_\text{ext}-Z(\omega)]/[Z_\text{ext}+Z(\omega)]$.
The other end is shunted by the capacitance of the small junction $C_J$, resulting in the reflection coefficient $[-1/(i\omega{C}_J)-Z(\omega)]/[-1/(i\omega{C}_J)+Z(\omega)]$. The Fabry-Perot round-trip resonance condition for a chain of $N\gg1$ junctions therefore reads as
\begin{equation}\label{eq:FabryPerot}
   \frac{Z_\text{ext}-Z(\omega)}{Z_\text{ext}+Z(\omega)}\,\frac{-1/(i\omega{C}_J)-Z(\omega)}{-1/(i\omega{C}_J)+Z(\omega)}\,e^{2iNk(\omega)} = 1.
\end{equation}
Since the chain's impedance $Z(\omega)\gg{Z}_\text{ext}=50\:\Omega$, the first factor can be set to $-1$ in the first approximation (the corresponding end of the chain being effectively grounded). Then Eq.~(\ref{eq:FabryPerot}) becomes
\begin{equation}
    Nk(\omega)+\arctan[\omega{C}_J{Z}(\omega)]=\pi(m-1/2),
    \label{eq:res_condition_simplified}
\end{equation}
whose real solutions $\omega_m$ for $m=1,2,\ldots$ are the resonant frequencies of the system. 
From the discrete resonant frequencies $\omega_m$, we can determine \begin{equation}\label{eq:nuomega}
\nu(\omega)=\frac{1}{\delta_m}\approx\frac{dm}{d\omega}=\frac{N}\pi\frac{dk}{d\omega}
+\frac1\pi\,\frac{d[\omega{C}_JZ(\omega)]/d\omega}{1+[\omega{C}_JZ(\omega)]^2}.
\end{equation}
The first $O(N)$ term is more important than the second $O(1)$ one in the long-chain limit.
Due to the nonlinear dependence of $k(\omega)$ on $\omega$ in Eq.~(\ref{eq:dispersion}), the mode frequencies of a long JJ chain are not equally spaced; $\delta_m$~shrinks by a factor of about $1.5$ at $\omega=\omega_\text{p}/2$ with respect to its low-frequency value. 

The coupling constants $\Lambda_m$ are given by Eq.~(\ref{eq:Lambdam2=}).
On the acoustic branch, $\omega_m\ll\omega_\text{p}$, we have $\delta_m\to\mathrm{const}$, $Z_\text{tot}(\omega_m)\to{Z}_0$, so $\Lambda_m\propto1/\sqrt{\omega_m}$. At the same time, the numerous modes near the plasma cutoff, $\omega_m\to\omega_\text{p}$, are effectively decoupled. Indeed, $\Re{Z_\text{tot}(\omega\to\omega_\text{p})}\approx\sqrt{1-\omega^2/\omega_\text{p}^2}/(\omega_\text{p}^2C_J^2Z_0)$, and $\delta_m/\omega_\text{p}\approx(\pi/N)\sqrt{C/C_\text{g}}(1-\omega_m^2/\omega_\text{p}^2)^{3/2}$, so $\Lambda_m\propto1-\omega_m^2/\omega_\text{p}^2$.

The losses $\kappa_m$ can be found as twice the imaginary parts of the complex solutions of the full Eq.~(\ref{eq:FabryPerot}) with $Z_\text{ext}$ included. Equivalently, since the losses $\kappa_m$ arise from the reflection at the boundary, we note that the energy stored at a given frequency $\omega$ is decreased by the intensity reflection coefficient $R(\omega)=|Z_\text{ext}-Z(\omega)|^2/{|Z_\text{ext}+Z(\omega)|^2}$ after a round trip time $t_R=2N dk/d\omega $ \cite{ismail_fabryperot_2016}. Therefore,
\begin{equation}
    \kappa_m = -\frac{1}{2N} \frac{d\omega}{dk} \ln R(\omega_m)\approx \frac{2\omega_\text{p}}{N}\sqrt{\frac{C}{C_\text{g}}} \,\frac{Z_{\text{ext}}}{Z_0} \left(1-\frac{\omega_m^2}{\omega_\text{p}^2}\right)^2.
\end{equation}
where we used $Z_0\gg Z_{\text{ext}}$ to obtain the second expression.
Keeping only the $O(N)$ term in $\nu(\omega)$ and adding phenomenologically an internal quality factor $Q_\text{int}$, we find
\begin{equation}
    \nu(\omega)\,\kappa(\omega)
    = \frac{2{Z}_\text{ext}}{\pi{Z}_0}\sqrt{1-\frac{\omega^2}{\omega_\text{p}^2}}
    +\frac{N\sqrt{C_\text{g}/C}}{\pi{Q}_\text{int}}
    \left(1-\frac{\omega^2}{\omega_\text{p}^2}\right)^{-3/2}.
\end{equation}
Typically, $Q_\text{int}$ is so large that the second term becomes important only in the close vicinity $\omega\to\omega_\text{p}$.

\begin{figure}
    \centering
    \includegraphics[width=0.48\textwidth]{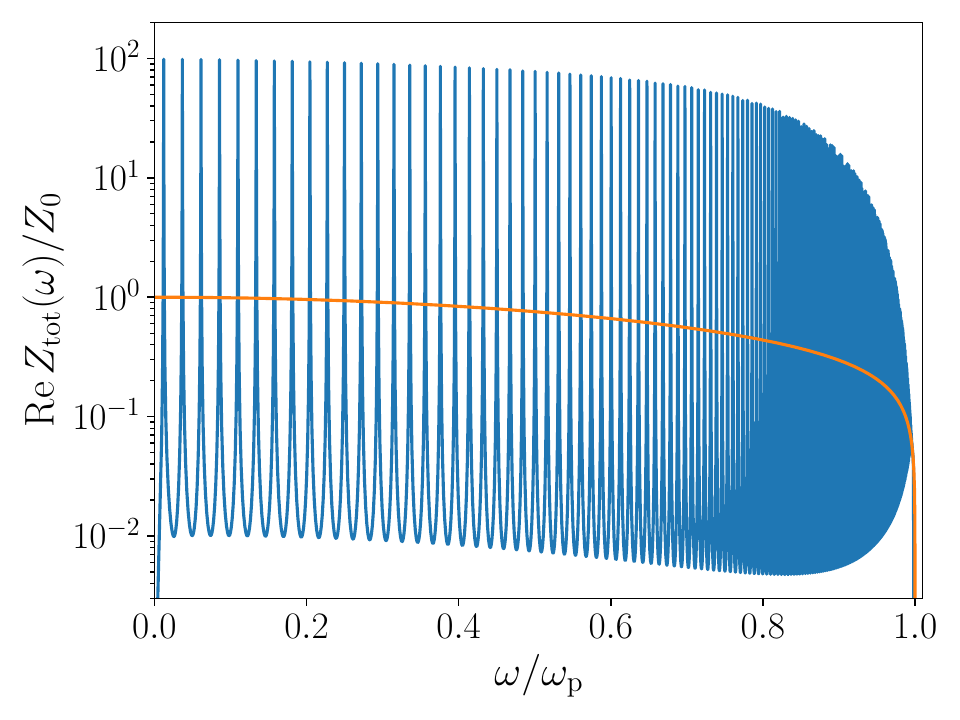}
    \caption{$\Re{Z_\text{tot}(\omega)}$ for an infinite JJ chain (smooth orange curve) and $\Re\mathcal{Z}_\text{tot}(\omega)$ for a chain of $N=5000$ junctions (spiky blue curve), all other parameters being the same as for Figs.~\ref{fig:selfconsistent_temp} and~\ref{fig:IVcurves}. }
    \label{fig:impedance_peaks}
\end{figure}

In Fig.~\ref{fig:impedance_peaks} we plot the resulting impedance of a finite chain, as given by Eq.~(\ref{eq:Z=Lambdak}). The frequency dependence consists of many sharp spikes, centered around the mode frequencies~$\omega_m$. If averaged over a small frequency interval exceeding $\delta_m$, it converges to the infinite-chain limit.

\section{Stability of the hot photonic state with respect to heating of quasiparticles}
\label{app:quasiparticles}

We found that the photonic degrees of freedom of the JJ chain can be overheated up to temperatures which may even exceed the critical temperature of the superconducting material. Can they affect (or even destroy) the superconductivity in the material? In the BCS theory, suppression of superconductivity occurs via proliferation of Bogolyubov quasiparticle excitations in the material. Assuming the experiment to have started at a low temperature, we can take the initial quasiparticle population to be very low. The frequencies of the photonic modes are well below the quasiparticle generation threshold $2\Delta$, so no new quasiparticles are produced directly by photons. 

The instability might develop if the few quasiparticles, that are already present in the material, absorb sufficient energy from the photons to go above the threshold $2\Delta$. Then they could produce new quasiparticles by breaking Cooper pairs via electron-electron collisions. Such quasiparticle heating via photon absorption is countered by cooling via phonon emission. We assume the phonon subsystem to be in equilibrium with a cryostat, so the phonon temperature (a few tens of milliKelvins) is smaller than the typical photon frequencies (the plasma frequency $\hbar\omega_\text{p}\sim2\pi\hbar\times20\:\mbox{GHz}\sim1\:\mbox{K}$), and can be set to zero.
Thus, we consider a single quasiparticle in the JJ chain and study the kinetic equation for its energy distribution function $f(\epsilon)$. An analytical solution for such kinetic equation was found in Ref.~\cite{catelani_nonequilibrium_2019} for a single JJ and a single photonic mode. It can be rather straightforwardly extended to the case of a JJ chain and with multiple photonic modes, as we show below.

We label the islands by an index~$\iota=0,\ldots,N$, and the quasiparticle states on each island by the orbital index~$p$ and spin projection~$\sigma$. Assuming all islands $\iota$ to have the same superconducting gap~$\Delta$, and focusing on quasiparticle energies close to the gap, we write the quasiparticle energies counted from the gap as
\begin{equation}
    \epsilon_{\iota,p}=\sqrt{\Delta^2+\xi_{\iota,p}^2}-\Delta\approx\frac{\xi_{\iota,p}^2}{2\Delta},
\end{equation}
where $\xi_{\iota,p}$ are the electron energies measured from the Fermi level in the normal state. They determine the normal density of states $\nu_0$ per spin projection or, equivalently, the inverse mean level spacing on each island,
\begin{equation}\label{eq:meanlevelspacing}
\nu_0V_\iota=\sum_p\delta(\xi_{\iota,p}),
\end{equation}
assumed to be energy-independent and proportional to the island volume~$V_\iota$. We also assume $\nu_0$ and $V_\iota$ to be the same for all islands.

The quasiparticles couple to the phase oscillations of the JJ chain when they tunnel across the JJs. Denoting the superconducting phase on each island~$\iota$ by $\hat\varphi_\iota$, we have the coupling Hamiltonian~\cite{catelani_quasiparticle_2011, catelani_relaxation_2011}:
\begin{align}
\hat{H}_\text{qp} ={}&{} \sum_{\iota,p,\sigma}
\epsilon_{\iota,p}\hat\gamma^\dagger_{\iota,p,\sigma}\hat\gamma_{\iota,p,\sigma}\nonumber\\
{}&{}+
\sum_{\iota=1}^{N-1}\sum_{p,\,p',\,\sigma} \left(i\mathcal{T}_{\iota+1/2,pp'}\hat{\gamma}_{\iota+1,p',\sigma}^{\dagger}\hat{\gamma}_{\iota,p,\sigma} + \text{h.c.}\right)\nonumber\\
{}&{}\qquad{}\times\sin \frac{\hat{\phi}_{\iota+1}-\hat\phi_\iota}{2},
\end{align}
where $\hat\gamma^\dagger_{\iota,p,\sigma}$ and $\hat\gamma_{\iota,p,\sigma}$, respectively, are the creation and annihilation operators for the quasiparticles. In a long chain of large JJs the phase difference between neighbouring islands is small, so we can approximate $\sin(\hat{\phi}_{\iota+1}-\hat\phi_\iota)/2\approx(\hat{\phi}_{\iota+1}-\hat\phi_\iota)/2$.
The tunneling matrix element $\mathcal{T}_{\iota+1/2,pp'}$ between the islands $\iota$ and $\iota+1$ is labeled by a half-integer $\iota+1/2$. It is assumed to be real and energy-independent on the relevant scale $\Delta$. These matrix elements determine the Josephson energy $E_{J}^\text{chain}$ of the large chain junctions (assumed to be identical) by the Ambegaokar-Baratoff relation~\cite{ambegaokar_tunneling_1963}:
\begin{equation}
\sum_{p,p'}\mathcal{T}_{\iota+1/2,pp'}^2
\delta(\xi_{\iota,p})\,\delta(\xi_{\iota+1,p'})
=\frac{E_{J}^\text{chain}}{\pi^2\Delta}.
\end{equation}

If a quasiparticle is initially on the level~$p$ of the island~$\iota$, it can absorb a photon while tunneling to a level $p'$ on the island $\iota+1$ or $\iota-1$. The rate of such transition is determined by Fermi's Golden Rule and can be written as
\begin{equation}
\Gamma_{\iota,p\to\iota+1,p'} = \frac{\mathcal{T}_{\iota,pp'}^2}{4\hbar^2}\,\mathcal{D}_{\iota+1/2}(\epsilon_{\iota,p}/\hbar-\epsilon_{\iota+1,p'}/\hbar),
\end{equation}
where $\mathcal{D}_{\iota+1/2}(\omega)$ is the phase difference correlator:
\begin{align}
\mathcal{D}_{\iota+1/2}(\omega)
={}&{}\int_{-\infty}^\infty{dt}\,e^{i\omega{t}}
\nonumber\\ {}&{}\times
\left\langle\left[\hat\phi_{\iota+1}(t)-\hat\phi_{\iota}(t)\right]\left[\hat\phi_{\iota+1}(0)-\hat\phi_{\iota}(0)\right]\right\rangle.
\end{align}
The transition rates give the following kinetic equation for the distribution function $f_\iota(\epsilon)$: 
\begin{widetext}\begin{align}
\frac{\partial{f}_\iota(\epsilon)}{\partial{t}} = {}&{} 
\frac{E_J^\text{chain}}{4\pi^2\hbar^2\Delta\nu_0V}
\int_0^\infty{d}\epsilon'\sqrt{\frac{2\Delta}{\epsilon'}}
\sum_\pm
\left[\mathcal{D}_{\iota\pm1/2}(\epsilon'/\hbar-\epsilon/\hbar)\,f_{\iota\pm{1}}(\epsilon')
-\mathcal{D}_{\iota\pm1/2}(\epsilon/\hbar-\epsilon'/\hbar)\,f_{\iota}(\epsilon)\right]
 + \text{St}_\text{n}f_\iota(\epsilon),
\end{align}\end{widetext}
where $\text{St}_\text{n}f_\iota(\epsilon)$ is the collision integral representing phonon emission by the quasiparticle, which we take the same as in Ref.~\cite{catelani_nonequilibrium_2019}.

Taking the mode spatial profiles for a $\lambda/4$ resonator with wave vectors $k(\omega_m)\equiv{k}_m=\pi(m-1/2)/N$,
\begin{equation}
\hat\phi_\iota(t) = \sum_{m=1}^N\Lambda_m\left(\hat{a}_me^{-i\omega_mt}+\hat{a}_m^\dagger e^{i\omega_mt}\right)\cos{k}_m\iota,
\end{equation}
we can evaluate the phase difference correlator as
\begin{align}
\left\langle\left[\hat\phi_{\iota+1}(t)-\hat\phi_{\iota}(t)\right]\left[\hat\phi_{\iota+1}(0)-\hat\phi_{\iota}(0)\right]\right\rangle \nonumber\\
{}=\sum_m4\Lambda_m^2\sin^2\frac{k_m}{2}\sin^2\left[k_m(\iota+1/2)\right]\nonumber\\
{}\times\left[(\bar{n}_m+1)\,e^{-i\omega_mt}+\bar{n}_me^{i\omega_mt}\right].
\end{align}
Noting that hot modes correspond to small wavevectors $k_m\ll1$ and assuming them to have the same temperature $T\gg \hbar\omega_m$ we approximate $\bar{n}(\omega_m) + 1 \approx \bar{n}(\omega_m)\approx{T}/(\hbar\omega_m)$. We also assume that the quasiparticle can be on any island with equal probability, $f_\iota(\epsilon)=f(\epsilon)$, so that we need the spatial average
\begin{equation}
\mathcal{D}(\omega)\equiv\sum_{\iota}\frac{\mathcal{D}_{\iota+1/2}(\omega)}N=
\frac{4e^2T}{\hbar^2}\,\frac{k^2(|\omega|)}{\omega^2}\,
\Re{Z_\text{tot}(\omega)},
\end{equation}
where the continuum limit $N\to\infty$ was taken.

As discussed in Ref.~\cite{catelani_nonequilibrium_2019}, two qualitatively different regimes can be identified, depending on the electron-phonon coupling strength. In the regime of weak overheating, the quasiparticle initially at low energy absorbs a photon, and then quickly emits a phonon without having a chance to absorb another photon; then the distribution function $f(\epsilon)$ mostly concentrated at low energies (of the order of the phonon temperature), with a weak single-photon replica. In the regime of strong overheating, the quasiparticle can absorb many photons before emitting a phonon.

Let us assume that this latter regime is realized, which corresponds to a pessimistic estimate. Then the typical photon energy $\hbar\omega$ is smaller than the characteristic width of the distribution function $f(\epsilon)$. Focusing on the high-energy tail of $f(\epsilon)$, we can (i)~approximate the quasiparticle-photon collision integral by a differential operator of the Fokker-Planck type, and (ii)~keep only the out-scattering term of the quasiparticle-photon collision integral. Then the kinetic equation takes the form~\cite{catelani_nonequilibrium_2019}
\begin{align}
\frac{\partial{f}(\epsilon)}{\partial{t}} =  {}&{}
\frac{E_J^\text{chain}}{\pi^2\hbar\nu_0V}
\int_0^\infty{d}\omega\,(\hbar\omega)^2\mathcal{D}(\omega)
\sqrt{\frac\epsilon{2\Delta}}\,\frac\partial{\partial\epsilon}\,\frac1\epsilon\,\frac{\partial{f}}{\partial\epsilon}\nonumber\\
{}&{} - \frac{128}{105}\,\frac{\epsilon^{7/2}}{\sqrt{2\Delta}T_c^3}\,\frac{f(\epsilon)}{\tau_0},
\label{eq:kinetic_Fokker_Planck}
\end{align}
where $T_c$ is the critical temperature of the superconductor, and $\tau_0$~is the typical timescale of electron-phonon scattering and is estimated to be $\tau_0\sim100\,\mathrm{ns}$ for thin aluminium films~\cite{moody_qp_relaxation}. 
The stationary solution of Eq.~(\ref{eq:kinetic_Fokker_Planck}) can be expressed in terms of the Airy function, and decays at high energies on the typical scale
\begin{equation}
    T_* = \left[ \frac{105}{128\pi^2}\,\frac{E_J^\text{chain}T_c^3}{\nu_0V(\hbar/\tau_0)} \int_0^\infty{d}\omega\,(\hbar\omega)^2\mathcal{D}(\omega)\right]^{1/6}.
\end{equation}
For an estimate, we take $\Re{Z_\text{tot}(\omega)}=Z_0$ in the interval $0<\omega<\omega_\text{p}$, $k^2(\omega)=(\omega/\omega_\text{p})^2(C_\text{g}/C)$, which gives $(T\hbar\omega_\text{p}/3)(C_\text{g}/C)(4e^2Z_0/\hbar)$ for the frequency integral. We note also that $E_J^\text{chain}=\hbar\omega_\text{p}\sqrt{C/C_\text{g}}\,\hbar/(4e^2Z_0)$.
Using $2\nu_0=2.15\times10^{47}\:\mbox{J}^{-7}m^{-3}=34.6\:\mbox{eV}^{-1}\mbox{nm}^{-3}$ for aluminium~\cite{anthore_density_2003} and taking the island volume $V=30\:\mbox{nm}\times2\:\mu\mbox{m}\times2\:\mu\mbox{m}\sim0.1\:\mu\mbox{m}^3$ we obtain the electronic level spacing on one island $1/(\nu_0V)\sim0.5\:\mbox{neV}$, and $\tau_0=100~\text{ns}$~\cite{moody_qp_relaxation} corresponds to $\hbar/\tau=6.6\:\text{neV}$.
Taking $\sqrt{C/C_\text{g}}=40$, $\omega_\text{p}=2\pi\times{20}\:\text{GHz}$, aluminum critical temperature $T_c=1.2\:\text{K}$, and mode temperature $T=3\,\hbar\omega_\text{p}$, we find $T_*/(\hbar\omega_\text{p})=0.25$. 

Since $T_*<\hbar\omega_\text{p}$, our initial pessimistic assumption of strong quasiparticle overheating is not valid. This means that the quasiparticle can reach energies $\epsilon\sim\hbar\omega_\text{p}$ at most. Thus, hot photons fail to heat up existing quasiparticles and to produce new ones, so the superconducting state in the material remains unaffected.

\section{Kinetic equation for the mode occupations}
\label{app:kinetic}

Most generally, the state of the environment can be described by the many-body density matrix in the basis of Fock states $|\{n_m\}\rangle$, specified by integer occupation numbers $n_m=0,1,\ldots$ of each mode~$m$. We now make several simplifying assumptions.
\renewcommand{\theenumi}{(\roman{enumi})}
\begin{enumerate}
\item
Assuming that the phases of different modes randomize, we take the density matrix to be diagonal in $\{n_m\}$.
\item
Assuming the mode frequency spectrum to be dense enough, we take the junction coupling to each individual mode to be small, $\Lambda_m\ll1$, so that a single Cooper pair tunneling event can change $n_m$ by $\pm1$ at most. This may be wrong for a few lowest modes, but we will check a posteriori that the contribution of these modes to the observables is regular.
\item
Assuming that in a single tunneling event only a small random fraction of all relevant $n_m$'s is changed, we  neglect correlations between different mode populations.
\end{enumerate}
As a result, we adopt the following ansatz for the many-body density matrix of the environment:
\begin{equation}\label{eq:diagonal_ansatz}
\rho_{\{n_m\},\{n_m'\}}=\prod_m\delta_{n_m,n_m'}p_m(n_m),
\end{equation}
where $p_m(n)$ is the mode-dependent distribution of occupation numbers, $\sum_{n=0}^\infty{p}_m(n)=1$.

\begin{widetext}
The mode occupations $n_m$ change with rates determined by Fermi Golden Rule. Let us choose some mode \qqq{$m_0$} and calculate the rate for transition between Fock states $|n_{m_0}\rangle$ and $|n_{m_0}'\rangle$, summing over all other modes \qqq{$m\neq{m_0}$}, as well as over forward/backward Cooper pair tunneling labeled by $\sigma=\pm$ (in the spirit of Refs.~\cite{ingold_charge_2005,hofheinz_bright_2011}):\begin{align}
\Gamma_{n_{m_0}\to{n}_{m_0}'}={}&{}\frac{E_J^2}{4\hbar}\sum_{\sigma=\pm}
\mathop{\sum\nolimits^\prime}\limits_{\{n_{m}\}}
\mathop{\sum\nolimits^\prime}\limits_{\{n_{m}'\}}
2\pi\delta\!\left(\sum_m(n_{m}-n_{m}')\hbar\omega_{m}+2eV\sigma\right)
\prod_m\left|\langle{n}_{m}'|e^{i\sigma\Lambda_{m}(\hat{a}_{m}+\hat{a}_{m}^\dagger)}|n_{m}\rangle\right|^2
\prod_{m\neq{m_0}}p_{m}(n_{m}){}\nonumber\\
={}&{}\frac{E_J^2}{4\hbar}\sum_{\sigma=\pm}
\int\limits_{-\infty}^\infty\frac{dt}{\hbar}\,e^{2ieV\sigma{t}/\hbar}
\mathop{\sum\nolimits^\prime}\limits_{\{n_{m}\}}
\mathop{\sum\nolimits^\prime}\limits_{\{n_{m}'\}}
\prod_m\left|\langle{n}_{m}'|e^{i\sigma\Lambda_{m}(\hat{a}_{m}+\hat{a}_{m}^\dagger)}|n_{m}\rangle\right|^2
e^{i(n_{m}-n_{m}')\omega_{m}t}\prod_{m\neq{m_0}}p_{m}(n_{m}),
\end{align}
where the primed sums are over all occupation numbers $\{n_{m}\}$ except~$n_{m_0}$ of the mode we are looking at. Neglecting terms $\Lambda_m^4$ and higher, we focus on the processes where $n'_{m}-n_{m}=0,\pm1$ and evaluate for an arbitrary $m\neq{m_0}$:
\begin{align}
&\sum_{n_{m},n_{m}'=0}^\infty\left|\langle{n}_{m}'|e^{i\sigma\Lambda_{m}(\hat{a}_{m}+\hat{a}_{m}^\dagger)}|n_{m'}\rangle\right|^2
e^{i(n_{m}-n_{m}')\omega_{m}t}p_{m}(n_{m}){}\nonumber\\
&\quad{}= \sum_{n_{m}=0}^\infty p_{m}(n_{m})\left[1-\Lambda_m^2(2n_{m}+1)+\Lambda_{m}^2(n_{m}+1)e^{-i\omega_{m}t}+\Lambda_{m}^2n_{m}e^{i\omega_{m}t}\right]+O(\Lambda_m^4) {}\nonumber\\
&\quad{}\approx1+\Lambda_{m}^2(\bar{n}_{m}+1)(e^{-i\omega_{m}t}-1)+\Lambda_{m}^2\bar{n}_{m}(e^{i\omega_{m}t}-1)\nonumber\\
&\quad{}\equiv1+J_{m}(t)\approx{e}^{J_{m}(t)},
\end{align}
where the average occupation $\bar{n}_{m}\equiv\sum_{n_{m}}n_{m}p_{m}(n_{m})$. 
Remarkably, the result of this first-order (in $\Lambda_m^2$) calculation is expressed in terms of the average occupation numbers, regardless of the exact form of the distribution $p_{m}(n_{m'})$. In higher orders, corresponding to emission or absorption of two or more photons in a given mode, higher moments of~$n_m$ would appear, which are sensitive to the specific distribution.
\end{widetext}

Assuming a large number of modes participating in the tunneling, we neglect the contribution of a single mode, and approximate 
\begin{equation}
\sum_{m\neq{m_0}}J_m(t)\approx\sum_mJ_m(t)\equiv{J}(t).
\end{equation}
Then we recover the definition~(\ref{eq:Joft}) with $\kappa_m\to0$ in Eq.~(\ref{eq:Z=Lambdak}) and unknown occupation numbers $\bar{n}_m=\bar{n}(\omega_m)$. Inclusion of finite damping rates $\kappa_m$ corresponds to Lorentzian broadening of the energy-conserving $\delta$~function in the Golden Rule, so the derivation remains consistent.
The transition rates can thus be expressed in terms of the function $P(E)$ as defined in Eq.~(\ref{eq:PofE}), again, depending on all occupations~$\{\bar{n}_m\}$:
\begin{align}
\Gamma_{n_m\to{n}_m\pm1} {}&{} = \frac{\pi{E}_J^2\Lambda_m^2}{2\hbar}\left(n_m+\frac12\pm\frac12\right)\sum_{\sigma=\pm}P(2eV\sigma\mp\hbar\omega_m)\nonumber\\
{}&{} \equiv \left(n_m+\frac12\pm\frac12\right)\Gamma_m^\pm.
\label{eq:bosonic_rates}
\end{align}

\begin{figure}
    \centering
    \includegraphics[width=0.48\textwidth]{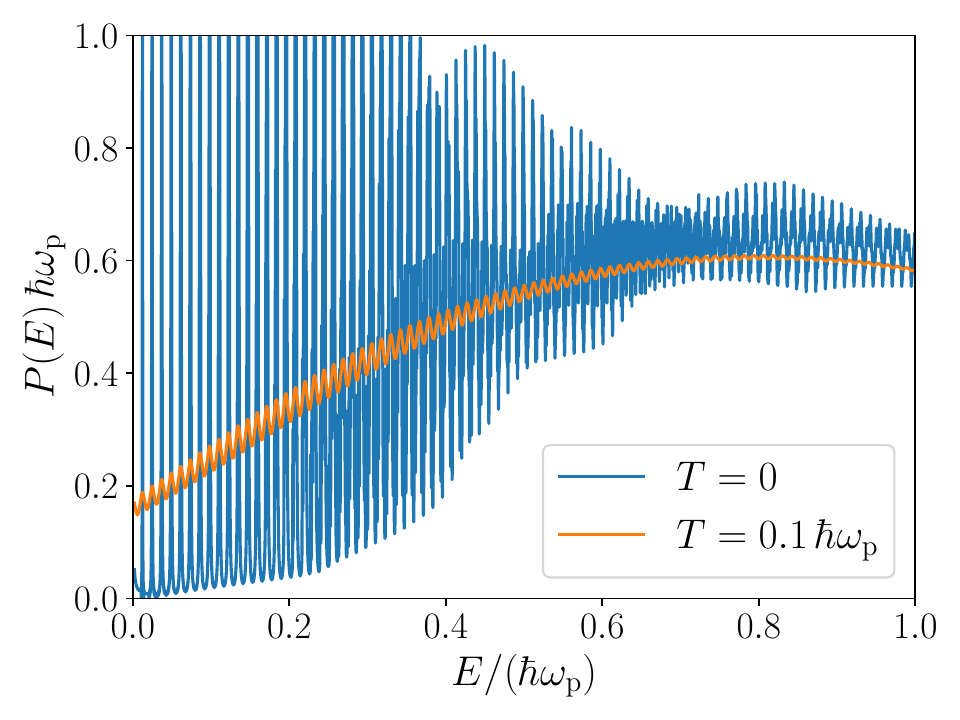}
    \caption{$P(E)$ calculated using ${Z}_\text{tot}(\omega)$ for a JJ chain of $N=5000$ junctions at $T=0$ (spiky blue curve) and $T=0.1\,\hbar\omega_\text{p}$ (weakly oscillating orange curve). All other parameters are the same as for Figs.~\ref{fig:selfconsistent_temp} and~\ref{fig:IVcurves}.}
    \label{fig:PofE_peaks}
\end{figure}

To verify the validity of the Golden Rule calculation, we compute $P(E)$ using Eqs.~(\ref{eq:PofE}) and~(\ref{eq:Joft}) with a thermal population $\bar{n}(\omega)=1/(e^{\hbar\omega/T}-1)$ and the finite-size impedance~$\mathcal{Z}_\text{tot}(\omega)$, given by Eq.~(\ref{eq:Z=Lambdak}) and shown in Fig.~\ref{fig:impedance_peaks} by the spiky curve. The result is shown in Fig.~\ref{fig:PofE_peaks}. While at zero temperature $P(E)$ approaches a smooth curve only at $E\sim\hbar\omega_\text{p}$ and higher, already at $T=0.1\,\hbar\omega_\text{p}$ $P(E)$ is almost smooth everywhere and close to the infinite-system limit (not shown, but checked). Since the temperature rises quite quickly with~$V$ (Fig.~\ref{fig:selfconsistent_temp}), our calculation based on the infinite-system impedance is consistent nearly everywhere, except at low voltages, $2eV\ll\hbar\omega_\text{p}$. The reason for this is that for a long JJ chain the single-photon frequency spacing~$\delta_m$ changes significantly over an interval of frequencies $\sim\omega_\text{p}$, as discussed in Appendix~\ref{app:param_computation}; then, the energies of multiphoton states which enter $P(E)$ have a much smaller spacing which becomes comparable with the escape rates $\kappa_m$, even though $\delta_m\gg\kappa_m$. For $\omega_m=m\omega_1$ or $\omega_m=(m-1/2)\omega_1$, there is a massive degeneracy of multiphoton states which invalidates the Golden Rule.

It is important to note that the aforementioned condition of smoothness for the $P(E)$ function, calculated using the impedance~$\mathcal{Z}_\text{tot}(\omega)$ from Eq.~(\ref{eq:Z=Lambdak}), does not involve $E_J$. Indeed, Eq.~(\ref{eq:Z=Lambdak}) includes mode broadening due to extrinsic mechanisms only, but not due to Cooper pair tunneling itself. Thus, it is just a sufficient condition. If we require the smoothness of $P(E)$ calculated with $\mathcal{Z}_\text{tot}(\omega)$ including the mode broadening $\Gamma_m^--\Gamma_m^+$, this would result in a weaker condition, corresponding to the condition of self-averaging of the self-energy calculated in the self-consistent Born approximation~\cite{basko_metal_2006}; indeed, $(\pi/8)E_J^2P(E)$ can be viewed as the imaginary part of the tunneling Cooper pair's self-energy. Here we do not perform this more complex self-consistent Born calculation, since the sufficient condition is already satisfied nearly everywhere.

Using the rates~(\ref{eq:bosonic_rates}), we can write a rate equation for $p_m(n_m)$. We also include the decay of excitations into the external circuit with the single-photon rate~$\kappa_m$, so that the transition rate for $n_m\to{n}_m-1$ involves the bosonic enhancement factor of~$n_m$:
\begin{align}
\frac{\partial{p}_m(n_m)}{\partial{t}}={}&{}
-\left(\Gamma_{n_m\to{n}_m+1}+\Gamma_{n_m\to{n}_m-1}+\kappa_mn_m\right)p_m(n_m){}\nonumber\\
{}&{} + \left[\Gamma_{n_m+1\to{n}_m}+\kappa_m(n_m+1)\right]p_m(n_m+1){}\nonumber\\
{}&{} + \Gamma_{n_m-1\to{n}_m}\,p_m(n_m-1).
\label{eq:rateeqn}
\end{align}
By virtue of the second line of Eq.~(\ref{eq:bosonic_rates}), the stationary solution of this equation has a thermal form:
\begin{align}\label{eq:thermal}
    p_m(n) = \frac{e^{-\zeta_m{n}}}{1-e^{-\zeta_m}},\quad
    \zeta_z\equiv\ln\frac{\Gamma_m^-+\kappa}{\Gamma_m^+}\equiv\frac{\hbar\omega_m}{T_m},
\end{align}
which defines the effective temperature $T_m$ of the mode~$m$ (not necessarily the same for all modes).
Multiplying Eq.~(\ref{eq:rateeqn}) by $n_m$ and summing over~$n_m$, we obtain the kinetic equation
\begin{equation}
\frac{\partial\bar{n}_m}{\partial{t}}=\Gamma_m^+(\bar{n}_m+1)-(\Gamma_m^-+\kappa_m)\bar{n}_m,
\label{eq:kinetic_m}
\end{equation}
which is a closed equation for the average occupations $\bar{n}_m$, since the rates~$\Gamma_m^\pm$ depend on the average occupations $\{\bar{n}_m\}$ via the $P(E)$, but not on higher moments of~$n_m$.
The stationary occupations are
\begin{equation}
\bar{n}_m=\frac{\Gamma_m^+}{\kappa_m+\Gamma_m^--\Gamma_m^+}
=\frac{1}{e^{\hbar\omega_m/T_m}-1}.
\end{equation}
Since $\Gamma_m^\pm$ depend on all $\{\bar{n}_m\}$, this gives a system of self-consistent equations for the average occupations.
To pass to the continuous frequency limit, we assume that $\bar{n}_m=\bar{n}(\omega_m)$, multiply the kinetic equation~(\ref{eq:kinetic_m}) by $\delta(\omega-\omega_m)\,\hbar\omega_m$ and sum over~$m$. This gives Eqs.~(\ref{eqs:kinetic}) where instead of the smooth functions $\nu(\omega)$ and $\Re{Z_\text{tot}(\omega)}$ we have, respectively, the spiky $\sum_m\delta(\omega-\omega_m)$ and $\Re{\mathcal{Z}_\text{tot}(\omega)}$ from Eq.~(\ref{eq:Z=Lambdak}) with $\kappa_m\to0$, which can also be written as a sum over $\delta(\omega-\omega_m)$. Finally, since for any function $\mathcal{F}(\omega)$ the equation $\sum_m\mathcal{F}(\omega_m)\,\delta(\omega-\omega_m) = 0$ is satisfied if $\mathcal{F}(\omega)=0$, we can effectively use Eqs.~(\ref{eqs:kinetic}) where only smooth functions of~$\omega$ enter.

To calculate the dc current in the stationary state, we first note that if we simply average the current operator
\begin{equation}
\hat{I} = \frac{2eE_J}{\hbar} \sin\left[\frac{2e}{\hbar}\,Vt - \sum_m\Lambda_m(\hat{a}_m+\hat{a}_m^\dagger)\right]
\end{equation}
over the diagonal density matrix~(\ref{eq:diagonal_ansatz}), its dc component vanishes, as for any density matrix diagonal in the Fock space of mode occupation numbers. To obtain off-diagonal matrix elements to the leading order in $E_J$, one should perturb the stationary diagonal density matrix $\hat\rho$ by the Josephson term $\hat{H}_J$ of the Hamiltonian [the first term of Eq.~(\ref{eq:Hamiltonian})]. Namely, we write the Hamiltonian as $\hat{H} = \hat{H}_0+\hat{H}_1(t)$, where 
\begin{equation}
    \hat{H}_0=\sum_m\hbar\omega_m\hat{a}_m^\dagger\hat{a}_m
    +\int_0^\infty{d\omega}\,    \hbar\omega\,\hat{b}^\dagger_\omega\hat{b}_\omega,
\end{equation}
and $\hat{H}_1(t)$ includes the time-dependent Josephson term and the coupling between the environment and the external circuit. Treating $\hat{H}_1$(t) as a perturbation, we obtain the correction to the diagonal density matrix:
\begin{align}\label{eq:rho1}
    \hat\rho_1(t) = \frac{i}\hbar\int\limits_{-\infty}^t{d}t'\,
    e^{-(i/\hbar)\hat{H}_0(t-t')}
    [\hat\rho,\hat{H}_1(t')]
    e^{(i/\hbar)\hat{H}_0(t-t')},
\end{align}
where $\hat{H}_1(t')$ is taken in the Schr\"odinger representation with the explicit time dependence determined by the voltage bias. 
A subtle point about the seemingly standard expression~(\ref{eq:rho1}) is that the stationary density matrix~$\hat\rho$ itself is a solution of the rate equations and thus contains arbitrarily high orders in $E_J^2$ and~$\kappa$, as seen explicitly from Eqs.~(\ref{eq:diagonal_ansatz}) and~(\ref{eq:thermal}). Solution of the rate equations is equivalent to resummation of an infinite series in the parameter $E_J^2/(\hbar^2\omega_\text{c}\kappa)$, which is necessary since $\kappa$ is small.  
The correction~(\ref{eq:rho1}) is small in the parameters $E_J/(\hbar\omega_\text{c})$, $\sqrt{\kappa/\omega_\text{c}}$ and thus is indeed perturbative.

This yields the first-order correction to the expectation value of the current, 
\begin{equation}
    I_1(t) = -\frac{i}{\hbar}\int_{-\infty}^t{d}t'\,\Tr\left\{\left[\hat{\tilde{I}}(t),\hat{\tilde{H}}_1(t')\right]\hat\rho\right\},
\end{equation}
where the tildes indicate  operators in the Heisenberg representation with respect to the Hamiltonian~$\hat{H}_0$, $\hat{\tilde{O}}(t)\equiv{e}^{(i/\hbar)\hat{H}_0t}\hat{O}{e}^{-(i/\hbar)\hat{H}_0t}$. To calculate the dc current, we need only the Josephson term in~$\hat{H}_1(t)$. For the same reason, when expanding the sine and cosine as difference and sum of exponentials, we keep only the terms proportional to $e^{\pm2ieV(t-t')/\hbar}$ and omit those oscillating as $e^{\pm2ieV(t+t')/\hbar}$.
Introducing
\begin{equation}
 \hat{\tilde\phi}(t)=
\sum_m\Lambda_m\left(\hat{a}_me^{-i\omega_mt}  + \hat{a}_me^{i\omega_mt}\right),
\end{equation}
and noting that half of the terms can be identically represented as a contribution from $t'>t$, we can write the dc current as
\begin{align}
I_\text{dc}  =
    \frac{eE_J^2}{2\hbar^2}
    \int_{-\infty}^\infty{d}\tau\left[e^{2ieV\tau/\hbar}\Tr\left\{e^{-i\hat{\tilde\phi}(\tau)}e^{i\hat{\tilde\phi}(0)}\hat\rho\right\} \right. \nonumber\\ 
    - \left.e^{-2ieV\tau/\hbar}\Tr\left\{e^{i\hat{\tilde\phi}(\tau)}e^{-i\hat{\tilde\phi}(0)}\hat\rho\right\}\right]
\end{align}
For the density matrix given by Eqs.~(\ref{eq:diagonal_ansatz}) and~(\ref{eq:thermal}), we can evaluate the correlators
\begin{equation}
    \Tr\left\{e^{-i\hat{\tilde\phi}(\tau)}e^{i\hat{\tilde\phi}(0)}\hat\rho\right\} = \Tr\left\{e^{i_m\hat{\tilde\phi}(\tau)}e^{-i\hat{\tilde\phi}(0)}\hat\rho\right\} = e^{J(t)},
\end{equation}
thus recovering Eqs.~(\ref{eqs:PofE}).

\section{Classical chaotic dynamics of the mode amplitudes}
\label{app:chaos}

\begin{figure*}
\includegraphics[width=0.45\textwidth]{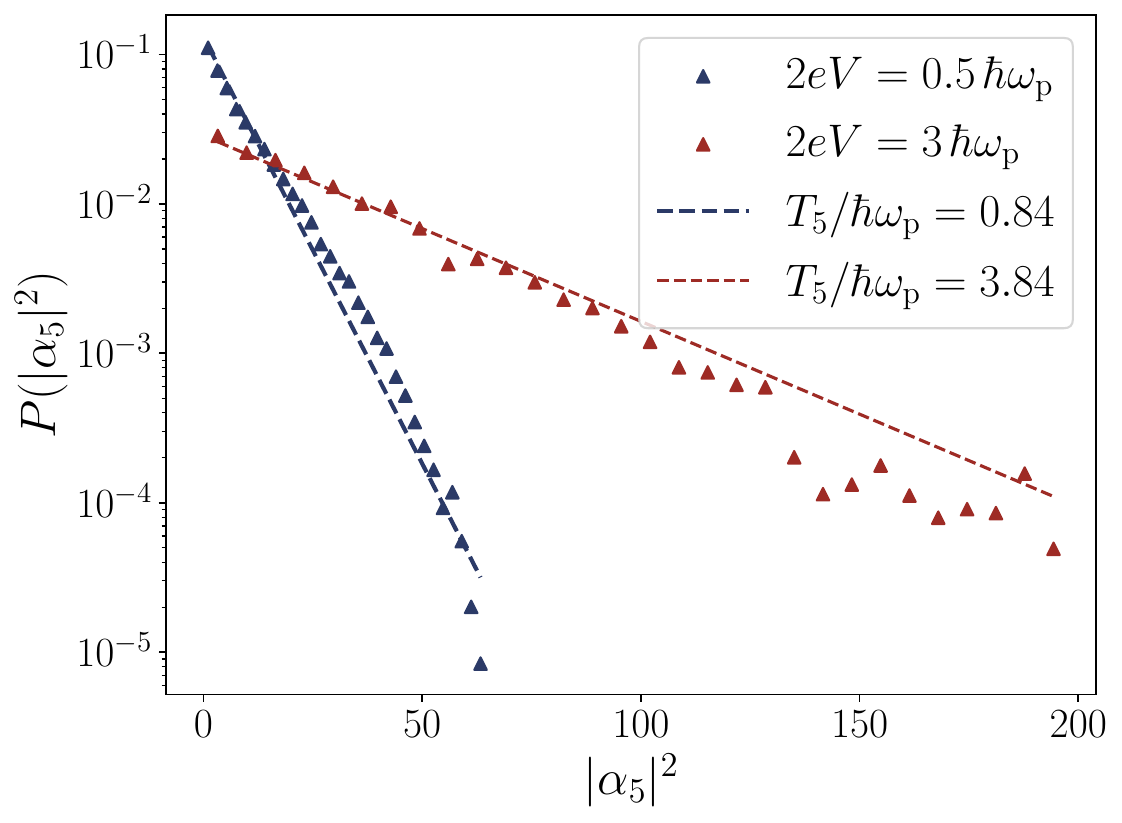}
    \hfill   \includegraphics[width=0.45\textwidth]{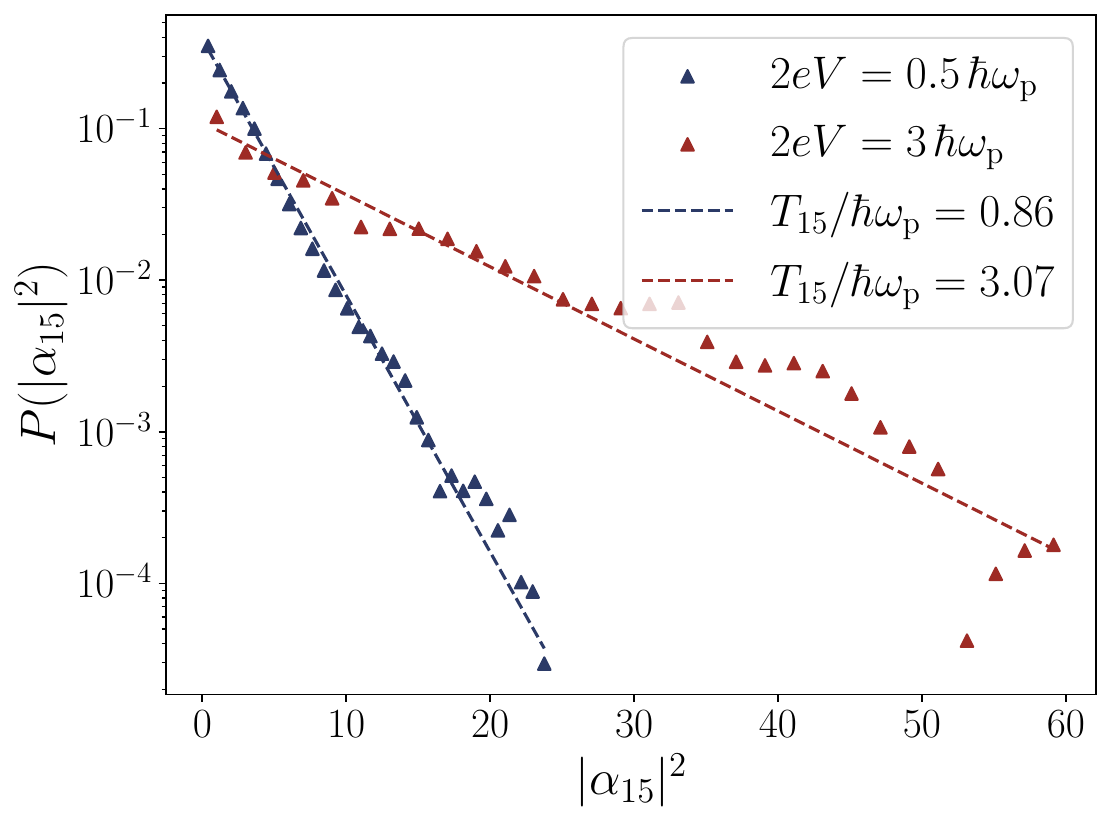}
    \caption{Probability distributions of $|\alpha_m|^2$ extracted by sampling the four trajectories in Fig.~\ref{fig:trajectories}. An exponential distribution $p_m(|\alpha_m|^2)=(\hbar\omega_m/{T_m}){e}^{-|\alpha_m|^2\hbar\omega_m/{T_m}}$ is also plotted (dotted lines) using the mode temperatures values $T_m$ from Fig.~\ref{fig:temperatures}.}
    \label{fig:radial_distribution}
\end{figure*}

\begin{figure*}
\includegraphics[width=0.8\textwidth]{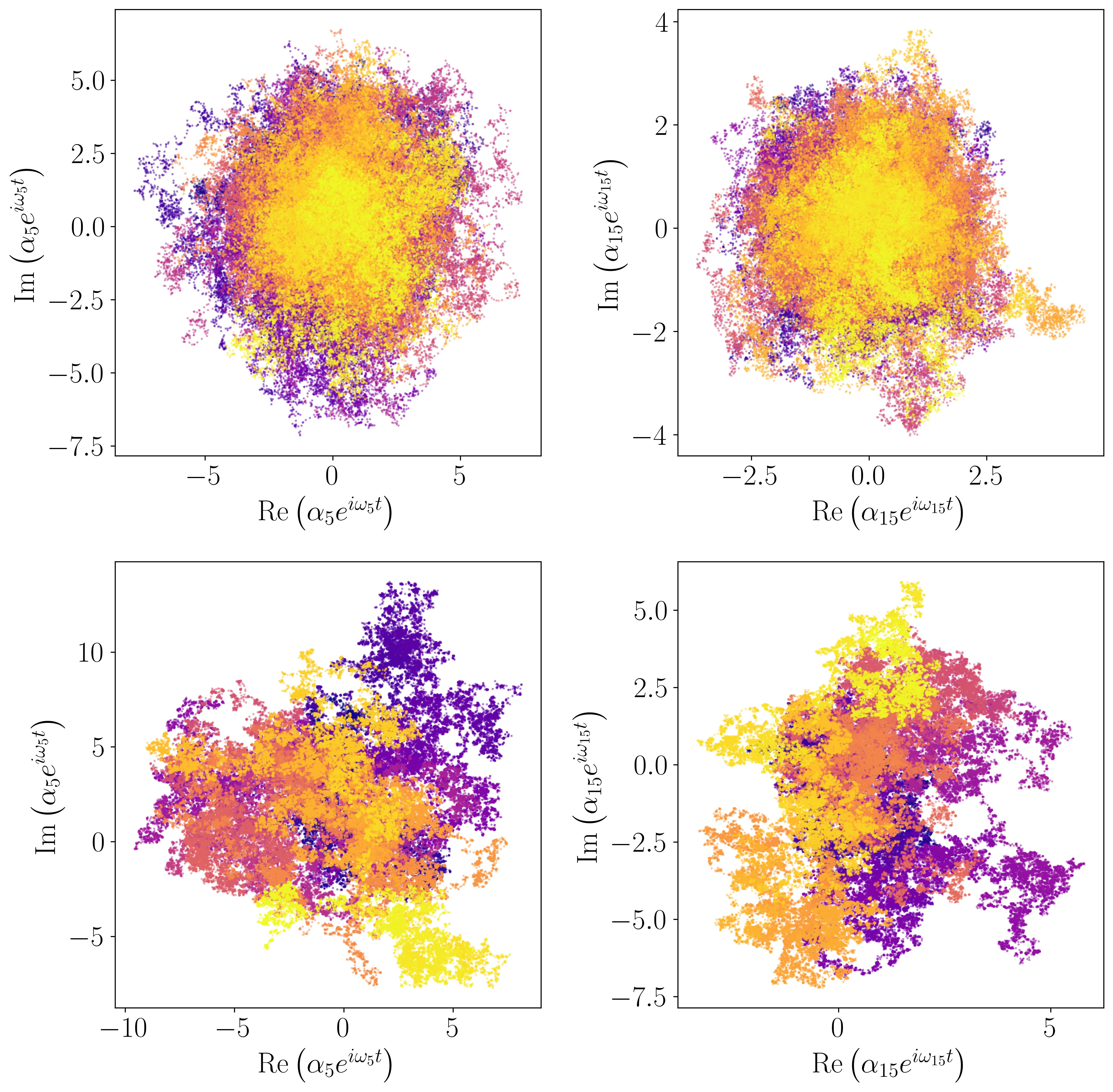}
    \caption{Trajectories of the slow envelopes $\alpha_m(t)\,e^{i\omega_mt}$ in the complex plane $(\Re\alpha,\Im\alpha)$ for the same system as in Figs.~\ref{fig:IVcurves} and~\ref{fig:temperatures} for two modes $m=5$ and $m=15$ (left and right column, respectively) at two voltages $2eV/(\hbar\omega_\text{p}) = 0.5$ and~$3.0$ (top and bottom row, respectively). The time is shown by color, from $\omega_\text{p} t=0$ (purple) to $\omega_\text{p} t=5\times 10^5$ (yellow).
}
    \label{fig:trajectories}
\end{figure*}

To see the chaotic character of the modes' classical dynamics described by Eqs.~(\ref{eqs:eom_modes}), we first look at the individual trajectories $\alpha_m(t)$ for a given voltage $V$, presented in Fig.~\ref{fig:trajectories}. The trajectories perform a random walk around the origin, which can be described by the Fokker-Planck limit ($n_m\gg1$) of Eq.~(\ref{eq:rateeqn}):
\begin{equation}
    \frac{\partial{p}}{\partial{t}} = \frac{\partial}{\partial{n}}
    \left[(\kappa_m+\Gamma_m^--\Gamma_m^+)np + \frac{\kappa_m+\Gamma_m^-+\Gamma_m^+}2\,n\,\frac{\partial{p}}{\partial{n}}\right].
\end{equation}
Its stationary solution corresponds to the Gaussian probability distribution for each~$\alpha_m$, $p_m(\alpha_m)\propto{e}^{-|\alpha_m|^2\hbar\omega_m/T_m}$. This, in turn, yields the Rayleigh-Jeans law for the average populations, $\overline{|\alpha_m|^2}=T_m/(\hbar\omega_m)$. The Gaussian statistics is confirmed by the numerical results for the radial distributions of $|\alpha_m|^2$, as presented in Fig.~\ref{fig:radial_distribution}. In Fig.~\ref{fig:trajectories} we show examples of individual mode trajectories from which this statistics is deduced.

\begin{figure*}    \includegraphics[width=0.45\textwidth]{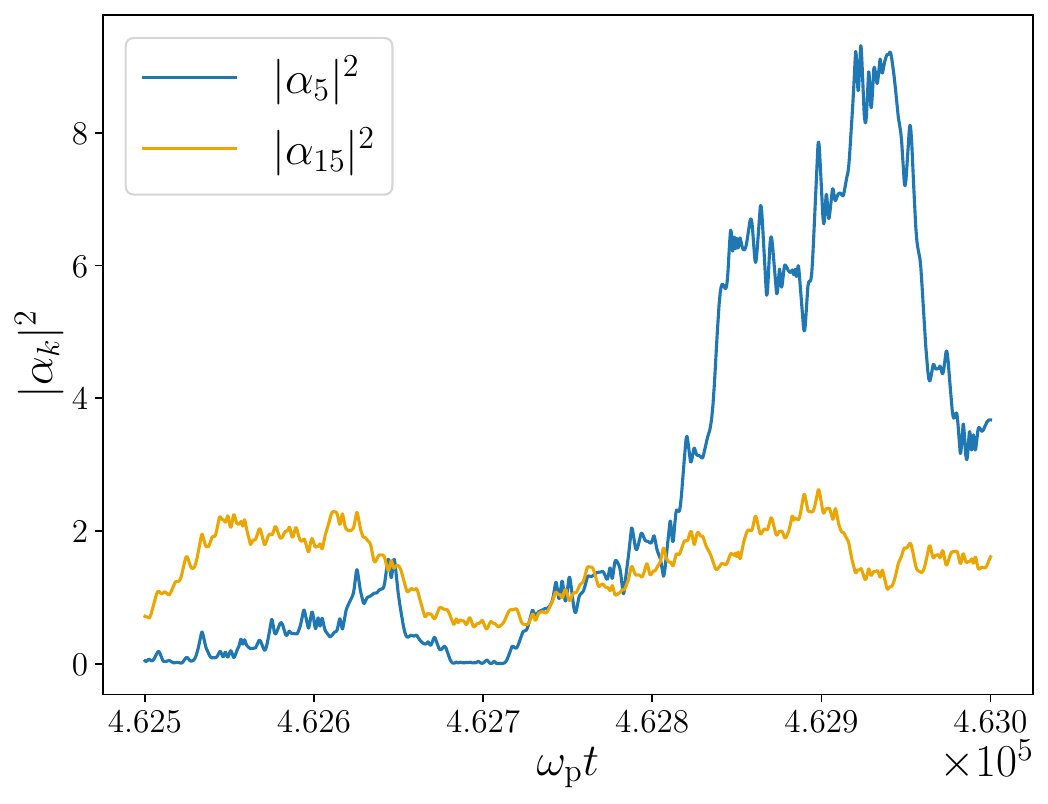}
    \hfill   \includegraphics[width=0.47\textwidth]{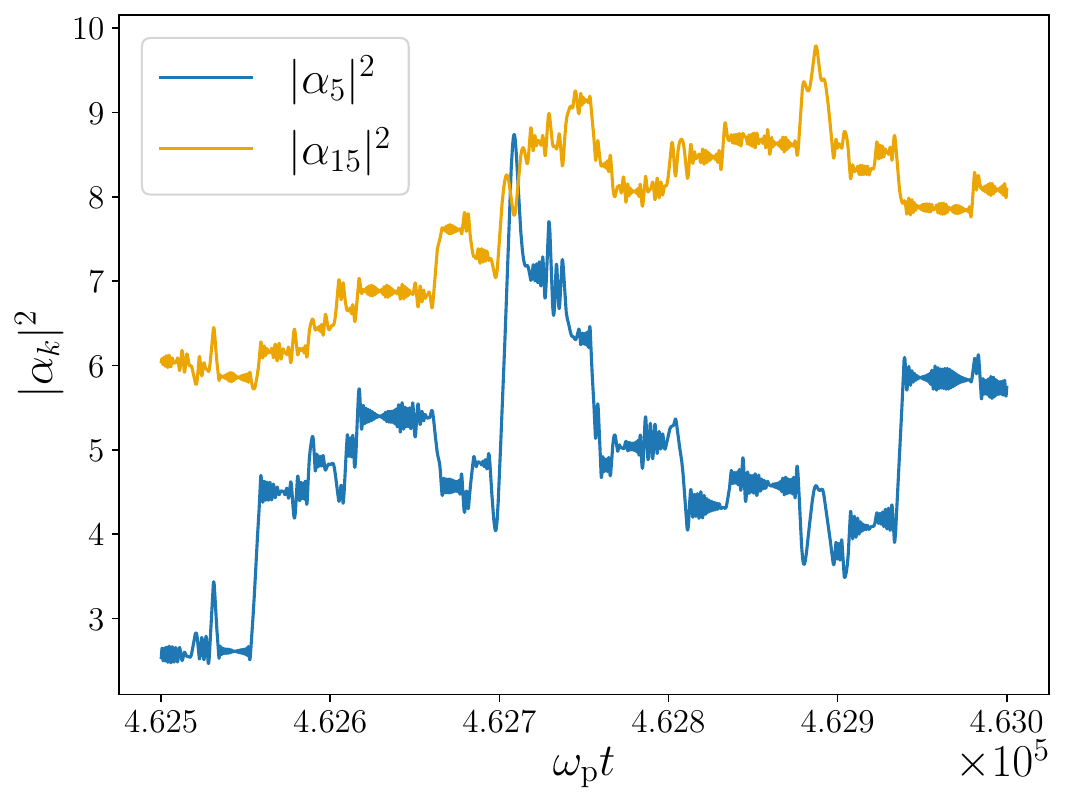}
    \caption{Evolution of $|\alpha_m(t)|^2$ for the two modes $m=5$ and $m=15$ at two voltages $2eV/(\hbar\omega_\text{p}) = 0.5$ and~$3.0$ (left and right panels, respectively).}
    \label{fig:short_times}
\end{figure*}

The modes' dynamics becomes chaotic on a time scale sufficiently long compared to the typical oscillation frequencies. 
Figure \ref{fig:short_times} shows evolution of the populations on a  timescale much shorter than that of Fig.~\ref{fig:trajectories}. We see that is consists of random quick jumps separated by long intervals of regular motion, and that populations of different modes jump simultaneously.

\section{Results for the capacitively shunted transmission line}
\label{app:transmission_line}

The expressions, given in this Appendix for a JJ chain, can also be used to describe a superconducting transmission line with linear dispersion. Namely, it is sufficient to take the limit $\omega_\text{p}\to\infty$, $C\to0$, $C\omega_\text{p}^2=\text{const}$, which gives $Z(\omega)=Z_0$ and $k(\omega)=\omega\sqrt{C_\text{g}/(C\omega_\text{p}^2)}$, coinciding with the acoustic branch dispersion of the JJ chain. 
For such a transmission line, the effective high-frequency cutoff is provided by the shunting capacitance of the small junction at frequencies $\omega\sim1/(Z_0C_J)$, since $\Lambda_m^2\propto\Re{Z_\text{tot}(\omega_m)}=Z_0/[1+(\omega_mC_JZ_0)^2]$.
The main difference with a JJ chain is that the mode frequency spacing $\delta_m$ is almost constant: as seen from Eq.~(\ref{eq:nuomega}), due to $dk/d\omega=\text{const}$, the relative change of~$\delta_m$ between $\omega\to0$ and $\omega\sim1/(Z_0C_J)$ is only $\sim1/N$. 

We now do the same calculations as described in the main text, but for a finite-length transmission line, whose parameters are chosen to match the low-frequency spectrum of the JJ chain. Namely, we take the same low-frequency impedance $Z_0=5\:\text{k}\Omega$, the small junction capacitance $C_J=2\:\text{f}F$ yields the soft cutoff frequency $\omega_\text{c}=1/(Z_0C_J)=2\pi\times15.9\:\text{GHz}$, close to the plasma frequency cutoff $\omega_\text{p}=20\:\text{GHz}$ of the studied JJ chain. The group velocity $v_\mathrm{g}=5\times 10^6\,\mathrm{m/s}$ is taken to match what can be achieved in high kinetic inductance materials~\cite{aiello_quantum_2022}, and we choose the transmission line length, $l=5\:\text{mm}$, in order to obtain the same mode spacing $\delta_m = \pi v_g/{l}$ at low frequencies, as for the JJ chain.
The Josephson energy of the small junction is taken as $E_J=0.2\,\hbar\omega_\text{c}$, the same fraction of the cutoff as before. Thus, we are comparing two systems whose low-energy properties are supposed to be similar.

\begin{figure*}
    \includegraphics[width=0.43\textwidth]{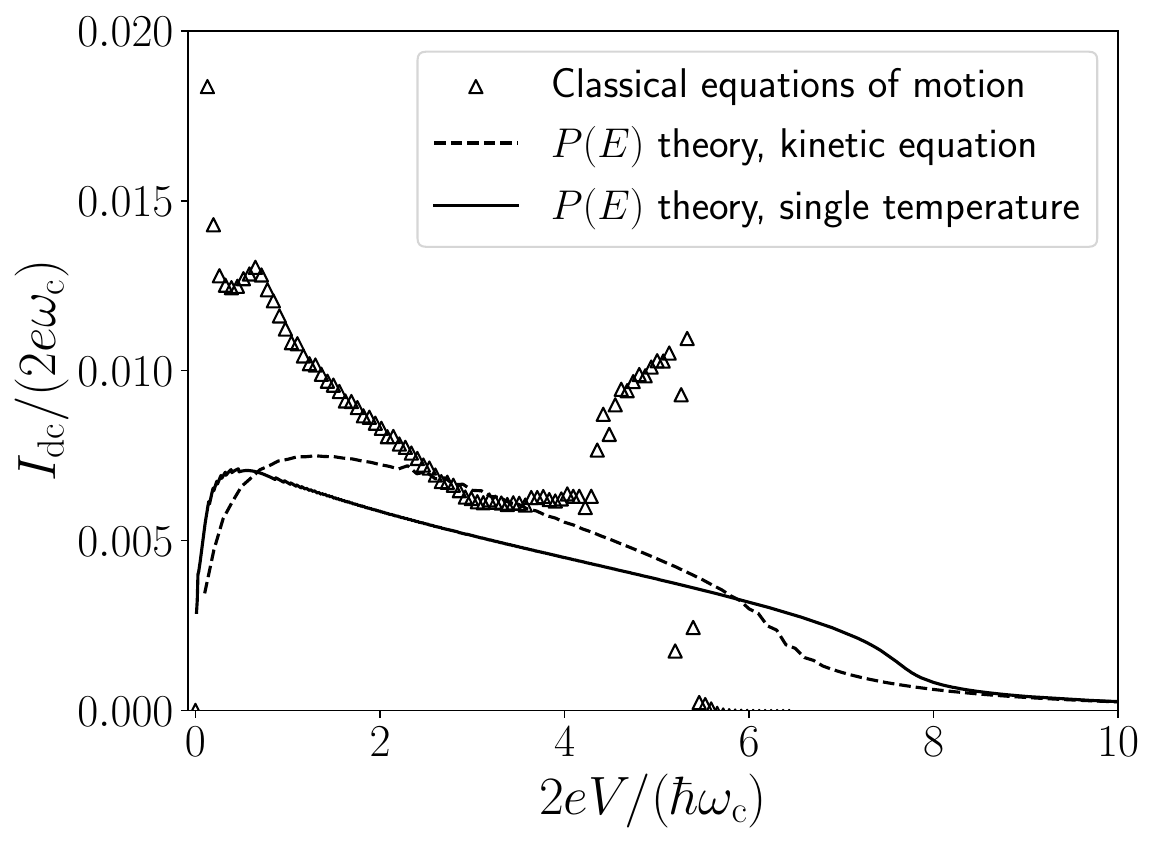}
    \hfill
    \includegraphics[width=0.53\textwidth]{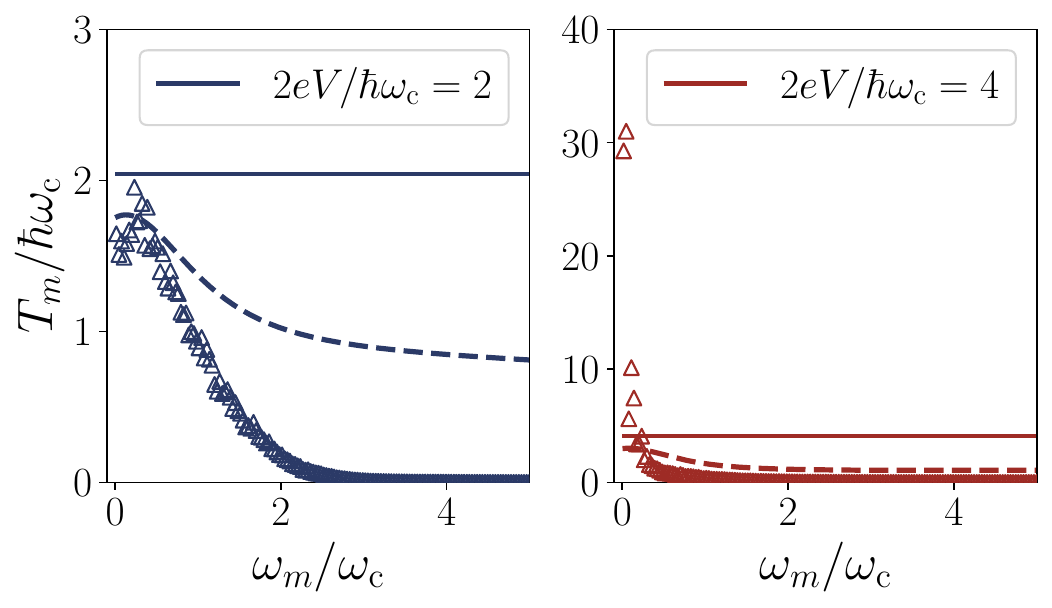}    
    \caption{$I-V$ curves (left panel) and mode temperatures (central and right panels) obtained from the three calculations described in the main text for a small junction coupled to a transmission line with constant group velocity $v_\text{g}=\partial \omega /{\partial k} = 5\times 10^6\,\mathrm{m/s}$, impedance $Z_0=5\,\mathrm{k\Omega}$ and length $l=5\,\mathrm{mm}$ yielding a frequency spacing $\delta_1 = \pi v_\mathrm{g} / {l} = 2\pi \times 0.5\,\mathrm{MHz}$. The small JJ parameters are $C_J=2\, \mathrm{fF}$ and $E_J=0.2\, \hbar\omega_\mathrm{c}$.}
    \label{fig:TL_ivAndTemps}
\end{figure*}

\begin{figure*}
    \includegraphics[width=0.45\textwidth]{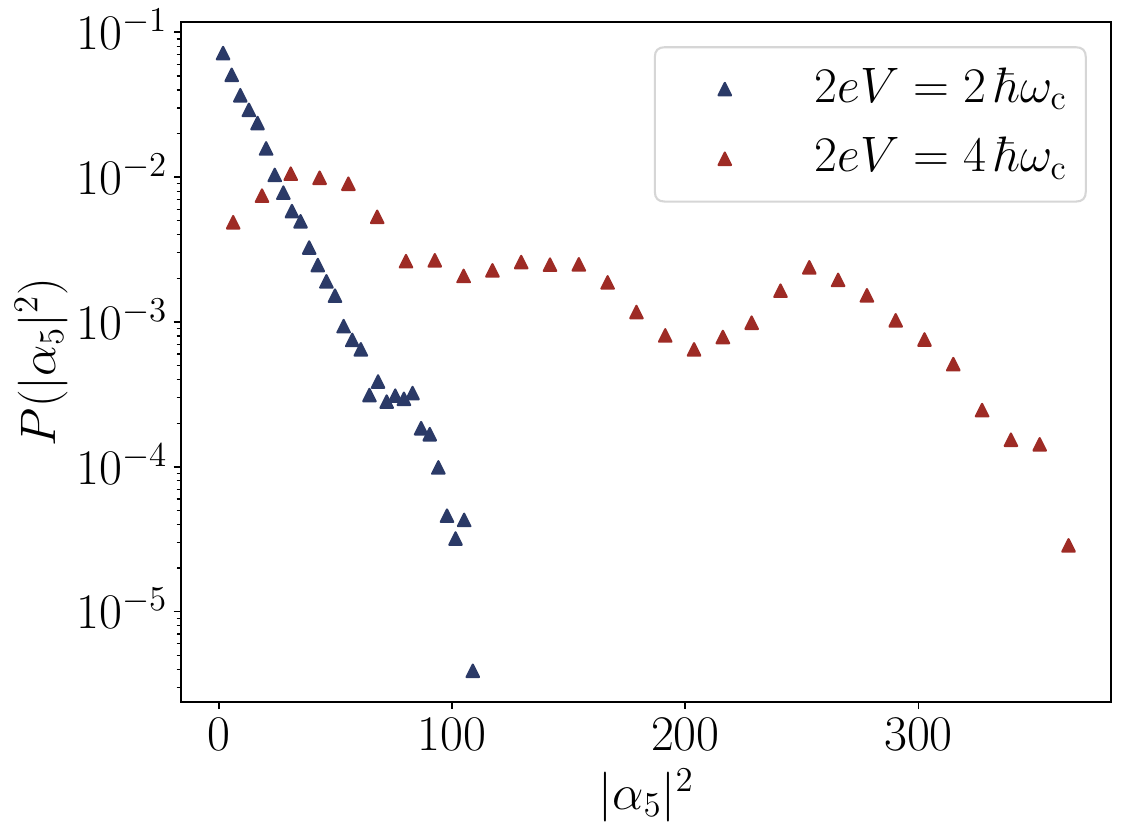}
    \hfill
    \includegraphics[width=0.45\textwidth]{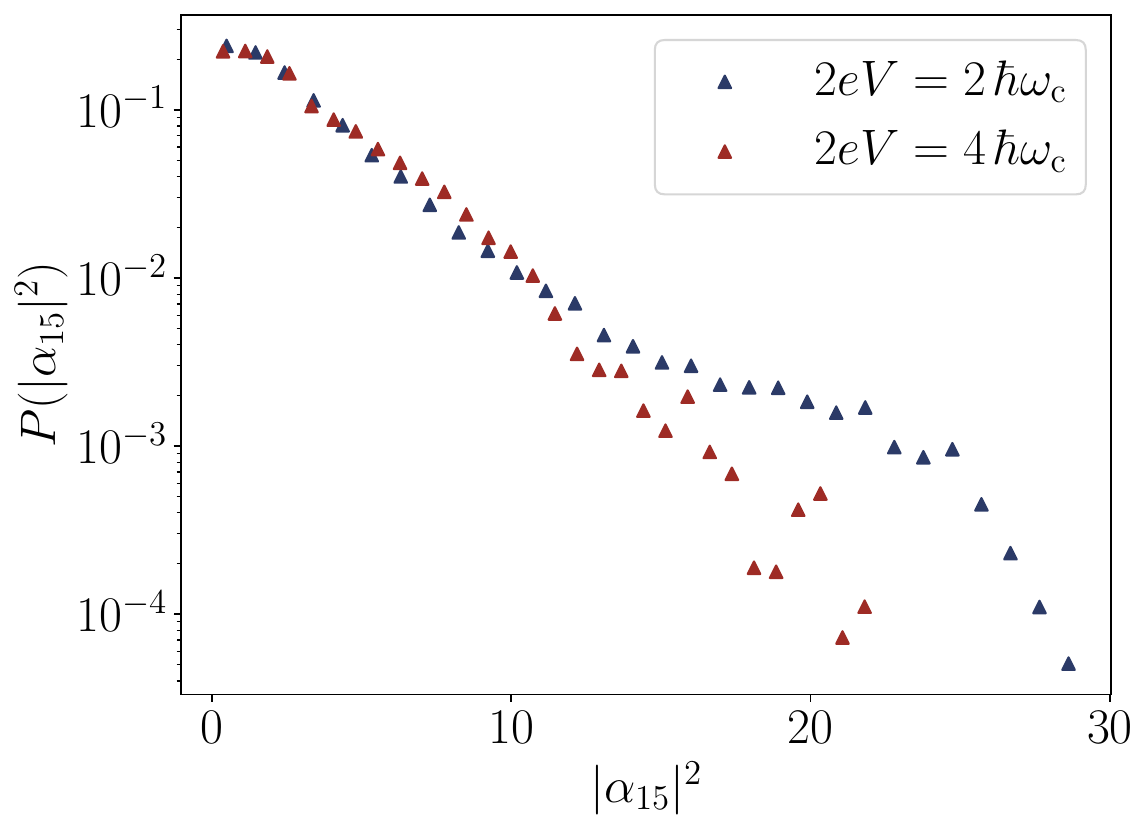}
    \caption{Probability distributions of $|\alpha_m|^2$ for the case of a dispersionless transmission line extracted in the same fashion as the one shown on Fig~\ref{fig:radial_distribution}. The statistics of $|\alpha_5|^2$ does not follow a centered Gaussian distribution at $2eV=4\,\hbar\omega_\mathrm{c}$.}
    \label{fig:radial_distribution_transmissionLine}
\end{figure*}

The resulting $I-V$ curves are shown in Fig.~\ref{fig:TL_ivAndTemps}, which shows much worse agreement between the different calculations than in Fig.~\ref{fig:IVcurves} for the JJ chain. In particular, the classical $I-V$ curve significantly deviates from both ones based on the $P(E)$ theory. For two values of the voltage bias, $2eV/(\hbar\omega_
\text{c})=2,\,4$ we also plot in Fig.~\ref{fig:TL_ivAndTemps} the average mode occupations, parametrized in terms of the corresponding effective temperatures, as it was done in Fig.~\ref{fig:temperatures} for the JJ chain. We observe that the classical calculation strongly favors high occupations of low-frequency modes, especially at higher voltages. Moreover, the distribution of the low-frequency mode occupations at this higher voltage becomes strongly non-thermal, as seen from Fig.~\ref{fig:radial_distribution_transmissionLine}; namely, the low values of the occupations become strongly suppressed.

These observations hint at a tendency for the system to develop a coherent component in its dynamics, which may be analogous to the Josephson laser~\cite{cassidy_demonstration_2017}. Its classical theory was developed in Ref.~\cite{simon_theory_2018}, where a coherent solution was explicitly constructed. Such coherent component is absent by construction in the $P(E)$ theory based calculations, which assume a diagonal density matrix in Fock space. On the other hand, the quantum kinetic $P(E)$ calculation keeps track of spontaneous emission into high-frequency modes, which thus takes away some energy from the drive and may prevent the energy from condensing into low-frequency modes.

We thus draw a preliminary conclusion that having a more regular equidistant spectrum of mode frequencies $\{\omega_m\}$ favors a tendency for the system to develop coherent dynamics. Quantifying this tendency and its relation to the frequency spectrum $\{\omega_m\}$ represents an interesting and difficult problem whose study is beyond the scope of the present paper.

\end{document}